\newcommand\p{\pi}

\newcommand\D{\Delta}

\newcommand{\diracslash}[1]{#1\llap{/\kern2pt}}

\newcommand{\be}{\begin{equation}}
\newcommand{\ee}{\end{equation}}
\newcommand{\bea}{\begin{eqnarray}}
\newcommand{\eea}{\end{eqnarray}}
\newcommand{\ba}[1]{\begin{array}{#1}}
\newcommand{\ea}{\end{array}}

\newcommand{\bt}{\begin{tabular}}
\newcommand{\et}{\end{tabular}}

\newcommand{\beas}{\begin{eqnarray*}}
\newcommand{\eeas}{\end{eqnarray*}}

\RequirePackage{lineno}

\documentclass[preprint,prd,aps,floats,nofootinbib,floatfix]
 {revtex4-2}

\DeclareSymbolFont{rsfs}{U}{rsfs}{m}{n}
\DeclareSymbolFontAlphabet{\mathrsfs}{rsfs}

\usepackage{amsmath}
\usepackage{graphicx}
\usepackage[mathscr]{eucal}
\usepackage{multirow}
\usepackage{graphicx,epstopdf}
\usepackage{graphicx}
\usepackage{subcaption}
\usepackage{cancel}

\usepackage{setspace}
\usepackage[utf8]{inputenc}
\usepackage{csquotes}
\usepackage[english]{babel}
\usepackage{hyperref}
\usepackage{upgreek}
\hypersetup{
    colorlinks=true,
    linkcolor=blue,
    filecolor=magenta,     
    citecolor=blue, 
    urlcolor=blue,
}
\usepackage{booktabs} 
\usepackage{cleveref}
\usepackage{caption}
\usepackage{rotating}

\begin{document}
\setstretch{1.5}
\title{Magnetic moments of decuplet baryons in isospin asymmetric magnetized strange matter} 
\author{Hrishika P}
\email{hrishikap2003@gmail.com}
\author{Manpreet Kaur}
\email{ranapreeti803@gmail.com}
\author{Suneel Dutt}
\email{dutts@nitj.ac.in}
\author{Harleen Dahiya}
\email{dahiyah@nitj.ac.in}

\author{Arvind Kumar}
\email{kumara@nitj.ac.in}
\affiliation{Department of Physics, Dr. B R Ambedkar National Institute of Technology Jalandhar, 
	Jalandhar -- 144008, Punjab, India}

\def\be{\begin{equation}}
\def\ee{\end{equation}}
\def\bearr{\begin{eqnarray}}
\def\eearr{\end{eqnarray}}
\def\zbf#1{{\bf {#1}}}
\def\bfm#1{\mbox{\boldmath $#1$}}
\def\hf{\frac{1}{2}}
\def\kp{\zbf k+\frac{\zbf q}{2}}
\def\km{-\zbf k+\frac{\zbf q}{2}}
\def\hwo{\hat\omega_1}
\def\hwt{\hat\omega_2}

\begin{abstract}
We investigate the in-medium masses and magnetic moments of decuplet baryons $(\Delta,\Sigma^*,\Xi^*,\Omega^-)$ in isospin asymmetric magnetized strange  matter at finite temperature within a unified chiral effective framework. Medium modifications of baryons are implemented using the chiral SU(3) quark mean-field (CQMF) model, where constituent quarks interact via scalar ($\sigma$, $\zeta$, $\delta$) and vector ($\omega$, $\rho$, $\phi$) meson fields considering the Dirac sea effects. The external magnetic field is incorporated through Landau quantization of charged particles together with anomalous magnetic moments (AMM) of baryons. The resulting in-medium mass of constituent quarks and decuplet baryons obtained from the CQMF model are subsequently employed as input to the chiral constituent quark model ($\chi$CQM) to evaluate magnetic moments of baryons. Contributions from valence quarks, sea quark spin polarizations, and orbital angular momentum of the quark sea are taken into account. Our results provide a systematic understanding of how dense, hot, and magnetized environments influence the magnetic properties of decuplet baryons.
\end{abstract}

\maketitle

\newpage

\section{Introduction}
\label{intro}The study of hadronic matter under extreme conditions of temperature, density, and magnetic field has become an essential direction in modern nuclear and particle physics. These studies offer a more profound insight into the non-perturbative elements of Quantum Chromodynamics (QCD), which governs the behavior of quarks and gluons.  It has been estimated that strong magnetic fields of the order of $eB\sim \,2m_\pi^2$ can be generated at Relativistic Heavy Ion Collider (RHIC), Brookhaven National Laboratory (BNL) and magnetic field as large as $eB\sim 15\,m_\pi^2$ is produced at the Large Hadron Collider (LHC) at CERN in noncentral HICs \cite{Kharzeev:2008npa,Skokov:2009mag,Voronyuk:2011prc,Bzdak:2012plb,Deng:2012prc,Bloczynski:2013plb,McLerran:2014npa,Tuchin:2016prc,Chen:2021npa}. These fields can significantly alter the internal structure and effective properties of hadrons by inducing anisotropy and modifying the QCD vacuum ~\cite{Kharzeev:2008npa,Fukushima:2008prd,Skokov:2009mag,Bali:2013jhep2}. 

Intense magnetic field induces many physical phenomena such as chiral magnetic effect (CME)~\cite{Kharzeev:2011prl,Kharzeev:2016ppnp,Kharzeev:2021natrev}, the
chiral magnetic wave (CMW)~\cite{Kharzeev:2016ppnp,Huang:2016rpp}, magnetic catalysis
(MC)~\cite{Gusynin:1996npb,Miransky_2002}, inverse magnetic catalysis (IMC)~\cite{Bali:2012prd}, the directed flow $v_1$ of $D^0$ mesons~\cite{Huang:2016rpp,Li:2020arnps,Huang:2023prc} and the
splitting of the spin polarization of the $\Lambda/\bar\Lambda$ hyperons. These phenomena have attracted considerable attention and have been extensively investigated through both theoretical and experimental studies~\cite{Fukushima:2019ppnp,Shovkovy:2022ws}. The phenomena of MC and IMC directly modify the chiral condensates, leading to significant changes in hadron masses and their interactions~\cite{PhysRevD.92.014038,Agasian:2008plb,Bali:2012jhep}. Furthermore, the observable changes in electromagnetic quantities such as quark anomalous magnetic moments, electric dipole moments, and spectral functions~\cite{Ferrer:2014prd,Fayazbakhsh:2014prd} influence the overall magnetic properties of composite hadrons \cite{Fraga:2024prd,PhysRevD.28.2881}. Over the past years, several theoretical frameworks have been developed to study hadronic properties in magnetized environments. These include relativistic mean-field (RMF) theories~\cite{Broderick:2002plb,Sinha:2013npa,Rabhi:2011prc}, Nambu–Jona-Lasinio (NJL) frameworks~\cite{Menezes:2009prc,Moreira:2021epja}, their polyakov extended versions (PNJL)~\cite{Wang:2022prd}, QCD sum rules (QCDSR)~\cite{Yue:2008prc}, confined isospin density dependent model~\cite{Dexheimer:2012epja,Chahal:2023prc,Chu:2018plb} etc. The inclusion of Dirac sea contributions has been shown to significantly influence the properties of magnetized matter by modifying scalar densities and vacuum structure~\cite{Menezes:2009prc2,Haber:2014prd,Mukherjee:2018prd,Aguirre:2016epja,Aguirre:2019prc}.


Magnetic moments and electromagnetic form factors of baryons are the primary observables that gives information about their internal quark structure~\cite{Savage2002,Leinweber1992,Alexandrou2009}. Magnetic moments of baryons have been investigated through different theoretical models, such as quark meson coupling (QMC) and modified QMC models~\cite{Tsushima:1998npa,Tsushima:2022ptep}, chiral perturbation theory ($\chi$PT)~\cite{Geng:2009,FloresMendieta:2009}, lattice quantum chromodynamics (LQCD)~\cite{Lee:2005ds}, chiral SU(3) quark mean-field (CQMF) model for isospin symmetric nuclear and strange matter, and asymmetric nuclear matter~\cite{Singh:2017cpc,Singh:2019epjp,Singh:2018epja,Singh:2020epjp}. Other approaches including chiral constituent quark model ($\chi$CQM) ~\cite{NSharma2010,Gridhar:2015prd} and statistical model ~\cite{Kaur2016EPJA} have also been employed to investigate baryon magnetic moments. In astrophysical systems, the anomalous magnetic moment of the neutron in dense magnetized matter has been investigated using the variational approach~\cite{Rezaei:2018ijmpe} and the influence of medium modified baryonic magnetic moments on neutron star properties in strong magnetic fields has been explored in Ref.~\cite{Ryu:2010prc}. Most recently, heavy baryon chiral perturbation theory (HB$\chi$PT) and covariant quark-diquark approaches have provided valuable frameworks for probing baryon structure beyond masses and magnetic moments. This has facilitated investigations of electromagnetic polarizabilities and gravitational form factors (GFFs), giving essential information about the dynamical and mechanical properties of hadrons~\cite{PhysRevD.105.096002,Wen2025EPJC}. 

The study of magnetic moments of decuplet baryons provides a sensitive probe of chiral symmetry breaking and medium induced modifications as these $J^P = \frac{3}{2}^+$ states receive contributions from valence quarks, polarized sea quarks, and orbital angular momentum of the quark sea~\cite{Manohar:1984georgi,Cheng:1997prl,Bernard:1995ijmpe}. The internal structure of $\Delta$ and other members of the decuplet baryons
have been theoretically explored in Refs.~\cite{PhysRevD.52.4099,PhysRevD.65.073017,PhysRevD.79.094025}. However, owing to the very short lifetimes of decuplet baryons, experimental information on their magnetic moments is limited, except for the $\Omega^-$ baryon, whose magnetic moment has been measured with reasonable precision. At present, limited experimental data exist primarily for the magnetic moments of the $\Delta$ and $\Omega^{-}$ baryons~\cite{PhysRevD.44.1962,PhysRevLett.67.804,PhysRevLett.74.3732,PhysRevLett.89.272001}, with $\Delta$ baryons measured with large uncertainties. Within the various frameworks discussed above, the magnetic moment of decuplet baryons are explored in 
Refs.~\cite{PhysRevD.48.4478,PhysRevD.57.1801,Tsushima:1998npa,Linde:1998PRD452,PhysRevD.67.114015,Gridhar:2015prd,Dahiya:2019owy,Singh:2018epja,Lee:2005ds,FloresMendieta:2009,Geng:2009}. These studies have been carried out in both free space and nuclear matter. More recently, the magnetic moments of decuplet baryons have been investigated within the CQMF framework in asymmetric strange hadronic matter at zero magnetic field, demonstrating the sensitivity of the decuplet baryons to the baryon density, isospin asymmetry, and strangeness fraction~\cite{Kumar:2023owb}. The effects of strong magnetic fields on decuplet baryon magnetic moments have also been explored in asymmetric magnetized nuclear matter, highlighting the significant role of the magnetic field in modifying the internal structure and electromagnetic properties of decuplet baryons~\cite{Dastidar:2026beh}.

Most existing studies are restricted to either nuclear matter or isospin symmetric strange matter, and a comprehensive analysis of decuplet baryon magnetic moment in isospin asymmetric strange matter under the influence of an external magnetic field including the contribution of Dirac sea to scalar densities is still lacking.
This motivates the present work where medium effects are incorporated through CQMF model, wherein quarks are treated as fundamental degrees of freedom and they interact via the exchange of scalar fields ($\sigma$, $\zeta$, $\delta$) and vector fields ($\omega$, $\rho$, $\phi$) along with dilaton field, $\chi$.
The external magnetic field is incorporated through Landau quantization of charged particles and anomalous magnetic moments of baryons, which collectively lead to further medium modifications of constituent quarks and baryons.
Using the consistently determined in-medium quark and baryon masses from the CQMF model, the magnetic moments of decuplet baryons are subsequently evaluated within the $\chi$CQM. In this framework, the valence quarks, the spin polarization of sea quarks and the orbital angular momentum of the quark sea contributes to the total magnetic moment of decuplet baryons. The presence of strange quarks introduces additional degrees of freedom that significantly influence the properties of hadrons in dense medium. Therefore, this study provides a unified description of decuplet baryon magnetic moments in isospin asymmetric magnetized strange matter by combining chiral mean field dynamics with the chiral constituent quark model. 

The structure of the paper is as follows: Section ~\ref{sec:cqmf} introduces the chiral SU(3) quark mean-field model with magnetic field, detailing the associated field equations and density formulas. The Sec.~\ref{sec:magnetic moment} focuses on the magnetic moments of decuplet baryons within a medium. Section ~\ref{sec:results} presents and discusses the numerical findings, while Sec.~\ref{sec:conclusion} concludes with a summary of the key results of this study.

\section{Chiral SU(3) quark mean field (CQMF) model}
\label{sec:cqmf}
In the present study medium-induced modifications of decuplet baryon are investigated using the CQMF framework, where the constituent quarks are confined inside baryons through an effective confining potential~\cite{Wang:2001npa,Papazoglou:1999prc}. The in-medium masses of constituent quarks arise from their interactions with scalar isoscalar non-strange $\sigma$, and strange $\zeta$, and the scalar isovector $\delta$ mesons within the medium. The trace anomaly is incorporated through the dilaton field, $\chi$. The effective Lagrangian density in this model is written as~\cite{Papazoglou:1999prc,Mishra2024Charmonium}
\begin{align}
 \mathcal{L}_{\text{eff}}= \mathcal{L}_{\text{chiral}}+\mathcal{L}_{\text{mag}} 
 \tag{1},
 \label{lagrangian}
\end{align}
where $\mathcal{L}_{\text{chiral}}$ is given by
\begin{equation}
\mathcal{L}_{\text{chiral}} \;=\; \mathcal{L}_{\text{q0}} + \mathcal{L}_{qm}
+ \mathcal{L}_{\text{VV}} + \mathcal{L}_{\Sigma \Sigma} + \mathcal{L}_{{\chi}SB} + \mathcal{L}_{\Delta m}+\mathcal{L}_{\text{c}}.
\tag{2}
\end{equation}
Here, $\mathcal{L}_{\text{q0}}= \bar{q}i\gamma^{\mu}\partial_{\mu}q$ represents the kinetic contribution of the massless quarks, while $\mathcal{L}_{qm}$ denotes the invariant quark-meson interaction sector~\cite{Ping:2001ctp,Singh:2017cpc}. 
The terms, $\mathcal{L}_{\Sigma\Sigma}$ and $\mathcal{L}_{VV}$
are the chiral invariant scalar and vector meson self interaction terms, respectively. Explicit chiral symmetry breaking is incorporated through $\mathcal{L}_{{\chi}SB}$ and the explicit strange quark mass contribution is introduced through $\mathcal{L}_{\Delta m}$. The last term, $\mathcal{L}_{c}$ represents the confinement of constituent quarks inside the baryons. The second term in Eq. (\ref{lagrangian}),
$\mathcal{L}_{\text{mag}}$ is added in order to include the external magnetic field effects. It describes the coupling of baryons to the external magnetic field, including the effects of anomalous magnetic moments (AMM)~\cite{Broderick:2000,Broderick:2002plb,Wei:2006,Mao:2003}, and is given by

\begin{equation}
\mathcal{L}_{\text{mag}} = -\bar\psi_i q_i \gamma_\mu A^\mu \psi_i - \tfrac12\kappa_i \bar\psi_i\sigma^{\mu\nu}F_{\mu\nu}\psi_i - \tfrac14 F^{\mu\nu}F_{\mu\nu}.
\tag{3}
\end{equation}
In the above equation, $\bar \psi_i$ and $q_i$ denote the wavefunction and electric charge associated with the $i^{th}$ baryon respectively. The second term, represents the tensorial interaction with the electromagnetic field, where $F_{\mu \nu}=\partial_\mu A_\nu - \partial_\nu A_\mu$, and $\sigma^{\mu \nu}=\frac{i}{2}[\gamma^\mu,\gamma^\nu]$. The quantity, $\kappa_i$ denotes the AMM of the baryon given by $\mu_i\mu_N$, where $\mu_i$ is the intrinsic magnetic moment and $\mu_N$  is the nuclear magneton. The inclusion of the electromagnetic interaction modifies the single particle energy spectrum and using these modified energy eigenvalues, the thermodynamic potential can be modified as
\begin{equation}
\Omega = \Omega_{\text{DS}} + \Omega_{\text{med}} - \mathcal{L}',
\tag{4}
\end{equation}
where $\Omega_{\text{DS}}$ and $\Omega_{\text{med}}$ denote the contributions of the
Dirac sea (DS) and the Fermi sea of baryons, respectively, to the total
thermodynamic potential and the last term $\mathcal{L}'$ is defined as
$\mathcal{L}' = \mathcal{L}_{VV} + \mathcal{L}_{\Sigma\Sigma}
+ \mathcal{L}_{\chi SB}$.
For spin-$\tfrac{1}{2}$ charged baryons ($i=p,\Sigma^{\pm},\Xi^{-}$), the contributions
from the Dirac sea to the thermodynamic potential are written as~\cite{Haber:2014prd,Aguirre:2016epja,Aguirre:2019prc}
\begin{equation}
\sum_i \Omega_{\text{DS}}^{\text{charged,}i}
= -\sum_i \frac{|q_i|B}{2\pi}\Bigg[\Bigg\{
\sum_{\bar \nu=0}^{\bar \nu_{max}^{(s=1)}}
\int_{-\infty}^{\infty}\frac{dk_z}{2\pi}\,
\epsilon^{\,i}_{k,\bar \nu,s}\Bigg\}
+
\Bigg\{
\sum_{\bar \nu=1}^{\bar \nu_{max}^{(s=-1)}}
\int_{-\infty}^{\infty}\frac{dk_z}{2\pi}\,
\epsilon^{\,i}_{k,\bar \nu,s}
\Bigg\}
\Bigg].
\tag{5}
\label{charged_ds}
\end{equation}
Here we can see that the continuous transverse momentum integration is replaced by a discrete sum over Landau levels $\bar \nu=0,1,2,...$$\bar \nu_{max}$, due to the quantization of charged particle orbits in the external magnetic field \cite{Landau:1965}. For neutral baryons ($i=n,\Lambda,\Sigma^0,\Xi^0$), which do not undergo Landau
quantization, the potential takes the form
\begin{equation}
\sum_{i} \Omega_{\text{DS}}^{\text{neutral},i}
= -\sum_i \sum_{s=\pm1}
\int \frac{d^3k}{(2\pi)^3}\,
\epsilon^{\,i}_{k,s}.
\tag{6}
\label{nuetral_ds}
\end{equation}
The medium contribution to thermodynamic potential for charged baryons is given by
\begin{align}
\sum_i \Omega_{\text{med}}^{\text{charged},i}
&= -T \sum_i \frac{|q_i|B}{2\pi}
\Bigg[
\sum_{\bar \nu=0}^{\bar \nu_{max}^{(s=1)}}
\int_{-\infty}^{\infty}\frac{dk_z}{2\pi}
\nonumber \\
&\quad \times \Bigg\{
\ln\!\left(1+e^{-\beta(\epsilon^{\,i}_{k,\bar \nu,s}-\nu_i^{*})}\right)
+ \ln\!\left(1+e^{-\beta(\epsilon^{\,i}_{k,\bar \nu,s}+\nu_i^{*})}\right)
\Bigg\}
\nonumber \\
&\quad + 
\sum_{\bar \nu=1}^{\bar \nu_{max}^{(s=-1)}}
\int_{-\infty}^{\infty}\frac{dk_z}{2\pi}
\Bigg\{
\ln\!\left(1+e^{-\beta(\epsilon^{\,i}_{k,\bar \nu,s}-\nu_i^{*})}\right)
+ \ln\!\left(1+e^{-\beta(\epsilon^{\,i}_{k,\bar \nu,s}+\nu_i^{*})}\right)
\Bigg\}
\Bigg],
\tag{7}
\label{charged_med}
\end{align}
and for neutral baryons, the contribution is given as
\begin{equation}
\sum_i \Omega_{\text{med}}^{\text{neutral},i}
= -T \sum_i \sum_{s=\pm1}
\int \frac{d^3k}{(2\pi)^3}
\Big[
\ln\!\left(1+e^{-\beta(\epsilon^{\,i}_{k,s}-\nu_i^{*})}\right)+ \ln\!\left(1+e^{-\beta(\epsilon^{\,i}_{k,s}+\nu_i^{*})}\right)
\Big],
\tag{8}
\label{neutral_med}
\end{equation}
where $\beta$ is the inverse of temperature $T$. The parameter $s=\pm1$ represents the spin up and spin down projections and $\nu^*_i$ is the effective chemical potential of baryons, which is a function of vector fields $\omega$, $\rho$ and $\phi$. In Eqs.  (\ref{charged_ds}) and  (\ref{charged_med}), $\epsilon^{\,i}_{k,\bar \nu,s}$ denotes the single particle energy spectra
of the $i^{th}$ charged baryon whereas $\epsilon^i_{k,s}$ in Eqs. (\ref{nuetral_ds}) and  (\ref{neutral_med}) denotes the energy spectra
of the $i^{th}$ neutral baryon in the presence of a magnetic field and are given as
\begin{equation}
\epsilon^{\,i}_{k,\bar \nu,s}
=
\sqrt{
k_z^2
+
\left(
\sqrt{2\bar \nu |q_i| B + M_i^{*2}} - s \kappa_i B
\right)^2
},\\
\label{Senergy1}
\tag{9}
\end{equation}
\text{and}
\begin{equation}
\epsilon^{\,i}_{k,s}
=
\sqrt{
k_z^2
+
\left(
\sqrt{k_x^2+k_y^2+M_i^{*2}} - s \kappa_i B
\right)^2
}.
\tag{10}
\label{Senergy2}
\end{equation}
In Eqs. (\ref{Senergy1}) and (\ref{Senergy2}), the components $k_x$ and $k_y$ represent the transverse momentum of the baryon, while $k_z$ denotes the longitudinal momentum component relative to the direction of the external magnetic field. Also, $M_i^*$ is the in-medium mass of the $i^{th}$ baryon defined as
\begin{equation}
M_i^{*}
= \sqrt{E_i^{*2} - \langle p_{i,\mathrm{cm}}^{2} \rangle},
\tag{11}
\label{baryonmass}
\end{equation}
where \(E_i^{*}\) denotes the effective in-medium energy of the baryon and
\(\langle p_{i,\mathrm{cm}}^{2} \rangle\) represents the correction arising from the
spurious center of mass motion of the constituent quarks and is evaluated as the sum of the average
squared momenta of the individual constituent quarks~\cite{Barik:2013prc,Barik:1985prd}
\begin{equation}
\langle p_{i,\mathrm{cm}}^{2} \rangle
= \sum_q \langle p_{\mathrm{cm}}^{*2} \rangle_q
= \sum_q
\frac{11 e_q^{*} + m_q^{*}}{6(3 e_q^{*} + m_q^{*})}
\left(e_q^{*2} - m_q^{*2}\right).\tag{12}
\end{equation}
Here $e_q^{*}$ is the effective energy of constituent quarks
and $m_q^*$ is the effective quark mass, which is given by
\begin{equation}
m_q^{*} = - g^q_{\sigma}\sigma - g_{\zeta}^q\zeta - g_{\delta }^qI^{3q} \delta +m_{q0},
\tag{13}
\end{equation} 
where the $g_\sigma^q$, $g_\zeta^q$ and $g_\delta^q$ are the coupling strength of quarks with the $\sigma$, $\zeta$ and $\delta$ fields and $m_{q0}$ is the bare quark mass contribution which is zero for non strange quarks and finite for strange quarks. The vacuum masses of constituent quarks are expressed in terms of vaccum values of fields, $\sigma_0$ and $\zeta_0$ as $m_u= m_d =
- \frac{g_{s}}{\sqrt{2}}\sigma_0=-g_\sigma^q\sigma_0$ and $m_s=-g_\zeta^s\zeta_0+m_1$ where $m_1=m_{q0}$. The numerical values of the coupling constants and vacuum fields are listed in Tables~\ref{params} and~\ref{tab:baryoncouplings}. The medium-induced modifications of the constituent quark masses lead to corresponding changes in the effective energies of the quarks, which are reflected in the effective energies of the baryons. 
Thus, the effective energy of the $i^{th}$ baryon can be expressed in
terms of the effective energies of its constituent quarks as
\begin{equation}
E_i^{*}
= \sum_q n_{qi}\, e_q^{*} + E_{i,\mathrm{spin}},
\tag{14}
\label{energyeqn}
\end{equation}
where \(n_{qi}\) denotes the number of $q$ quarks in the $i^{th}$ baryon.
The term \(E_{i,\mathrm{spin}}\) accounts for the spin-spin interaction among the
constituent quarks and serves as a correction to the baryon energy. The isospin asymmetry parameter is defined as $I_a = -\sum_i \frac{I^{3i}\rho^i_v}{\rho_B}$, where $I^{3i}$ represents the third component of isospin and $\rho_B$ is the total baryonic density of the medium.  The strangeness fraction is given by
$f_s=\sum_i \frac{|s_i|\rho^i_v}{\rho_B}$, where $|s_i|$ represents the number of strange
quarks in the $i^{th}$ baryon. To determine the equilibrium state of the system for given values of $\rho_B$, $I_a$, $f_s$,  $T$, and magnetic field $eB$, the total thermodynamic potential is minimized with respect to the mesonic fields $(\sigma,\zeta,\delta,\chi,\omega,\rho,\phi)$~\cite{Kumari:2020mci,Kumar:2019cpc,Kumar:2019epjc} such that,
\begin{equation}
\frac{\partial \Omega}{\partial \sigma}
=
\frac{\partial \Omega}{\partial \zeta}
=
\frac{\partial \Omega}{\partial \delta}
=
\frac{\partial \Omega}{\partial \chi}
=
\frac{\partial \Omega}{\partial \omega}
=
\frac{\partial \Omega}{\partial \rho}
=
\frac{\partial \Omega}{\partial \phi}
= 0 .
\tag{15}
\end{equation}
The minimization conditions lead to the following nonlinear coupled mean-field equations
\setcounter{equation}{15}
\begin{eqnarray}
\frac{\partial \Omega}{\partial \sigma} &=& k_0 \chi^2 \sigma 
- 4k_1 (\sigma^2 + \zeta^2 + \delta^2)\sigma 
- 2k_2 (\sigma^3 + 3\sigma \delta^2) 
- 2k_3 \chi \sigma \zeta \nonumber -\\&& \frac{d}{3}\chi^4 \left(\frac{2\sigma}{\sigma^2 - \delta^2}\right)+ \left(\frac{\chi}{\chi_0}\right)^2 m_\pi^2 f_\pi 
- \left(\frac{\chi}{\chi_0}\right)^2 m_\omega \omega^2 \frac{\partial m_\omega}{\partial \sigma} \nonumber - \\&&\left(\frac{\chi}{\chi_0}\right)^2 m_\rho \rho^2 \frac{\partial m_\rho}{\partial \sigma}
- \sum_i g_\sigma^i \rho_s^i=0,
\end{eqnarray}
\begin{eqnarray}
\frac{\partial \Omega}{\partial \zeta} &=& k_0 \chi^2 \zeta 
- 4k_1 (\sigma^2 + \zeta^2 + \delta^2)\zeta 
- 4k_2 \zeta^3 
- k_3 \chi (\sigma^2 - \delta^2) \nonumber-\\&& \frac{d}{3}\frac{\chi^4}{\zeta}
+ \left(\frac{\chi}{\chi_0}\right)^2 \left[
\sqrt{2} m_K^2 f_K - \frac{1}{\sqrt{2}} m_\pi^2 f_\pi
\right] \nonumber  - \left(\frac{\chi}{\chi_0}\right)^2 m_\phi \phi^2 \frac{\partial m_\phi}{\partial \zeta}
 \\&&-\sum_i g_\zeta^i \rho_s^i=0,
\end{eqnarray}
\begin{eqnarray}
\frac{\partial \Omega}{\partial \delta} &=& k_0 \chi^2 \delta 
- 4k_1 (\sigma^2 + \zeta^2 + \delta^2)\delta 
- 2k_2 (\delta^3 + 3\sigma^2 \delta) 
+ 2k_3 \chi \delta \zeta \nonumber \\
&& + \frac{2}{3} d \chi^4 \left(\frac{\delta}{\sigma^2 - \delta^2}\right)
- \sum_i g_\delta^i I^{3i}\rho_s^i =0,
\end{eqnarray}
\begin{eqnarray}
\frac{\partial \Omega}{\partial \chi} &=& k_0 \chi (\sigma^2 + \zeta^2 + \delta^2)
- k_3 (\sigma^2 - \delta^2)\zeta 
+ \chi^3 \left[1 + \ln\left(\frac{\chi^4}{\chi_0^4}\right)\right] 
+ (4k_4 - d)\chi^3 \nonumber \\
&& - \frac{4}{3} d \chi^3 \ln\left(
\frac{(\sigma^2 - \delta^2)\zeta}{\sigma_0^2 \zeta_0}
\left(\frac{\chi}{\chi_0}\right)^3
\right) \nonumber + \frac{2\chi}{\chi_0^2} \left[
m_\pi^2 f_\pi \sigma 
+ \left(\sqrt{2} m_K^2 f_K - \frac{1}{\sqrt{2}} m_\pi^2 f_\pi \right)\zeta
\right] \nonumber \\
&& - \frac{\chi}{\chi_0^2}(m_\omega^2 \omega^2 + m_\rho^2 \rho^2+m_\phi^2\phi^2)=0,
\end{eqnarray}
\begin{eqnarray}
\frac{\partial \Omega}{\partial \omega} &=& \frac{\chi^2}{\chi_0^2} m_\omega^2 \omega 
+ 4g_4 \omega^3 
+ 12g_4 \omega \rho^2 
- \sum_i g_\omega^i \rho_v^i =0,
\end{eqnarray}
\begin{eqnarray}
\frac{\partial \Omega}{\partial \rho} &=& \frac{\chi^2}{\chi_0^2} m_\rho^2 \rho 
+ 4g_4 \rho^3 
+ 12g_4 \omega^2 \rho 
- \sum_i g_\rho^i I^{3i}\rho_v^i =0,
\end{eqnarray}
\begin{eqnarray}
\frac{\partial \Omega}{\partial \phi} &=& \frac{\chi^2}{\chi_0^2} m_\phi^2 \phi 
+ 8g_4 \phi^3 
- \sum g_\phi^i \rho_v^i =0.
\end{eqnarray}
The number density of the $i^{th}$ baryon, $\rho_v^i$ is obtained from the medium
contribution to the thermodynamic potential. For charged baryons, it is given by
\begin{equation}
\begin{aligned}
\rho_v^i
&= \frac{|q_i|B}{2\pi^2}
\Bigg[
\sum_{\bar \nu=0}^{\bar \nu_{max}^{(s=1)}}
\int_{0}^{\infty} dk_z
\Bigg\{
\frac{1}{1+e^{\beta(\epsilon^{i}_{k,\bar \nu,s}-\nu_i^{*})}}
-
\frac{1}{1+e^{\beta(\epsilon^{i}_{k,\bar \nu,s}+\nu_i^{*})}}
\Bigg\}
\nonumber\\
&\quad+
\sum_{\bar \nu=1}^{\bar \nu_{max}^{(s=-1)}}
\int_{0}^{\infty} dk_z
\Bigg\{
\frac{1}{1+e^{\beta(\epsilon^{i}_{k,\bar \nu,s}-\nu_i^{*})}}
-
\frac{1}{1+e^{\beta(\epsilon^{i}_{k,\bar \nu,s}+\nu_i^{*})}}
\Bigg\}
\Bigg].
\end{aligned}
\end{equation}
For neutral baryons the corresponding expression is given as 
\begin{equation}
\rho_v^i
= \sum_{s=\pm1}
\int \frac{d^3k}{(2\pi)^3}
\left[
\frac{1}{1+e^{\beta(\epsilon^{i}_{k,s}-\nu_i^{*})}}
-
\frac{1}{1+e^{\beta(\epsilon^{i}_{k,s}+\nu_i^{*})}}
\right]. 
\end{equation}
By neglecting the Dirac sea contribution to the thermodynamic potential,
the medium contribution to the scalar density $\rho_s^i$ for charged baryons is given by
\begin{equation}
\begin{aligned}
 \rho_{s,\mathrm{med}}^{i}
&= \frac{|q_i|B}{2\pi^2}\Bigg[\sum_{\bar \nu=0}^{\bar \nu_{max}^{(s=1)}}M_i^{*}
\int_{0}^{\infty} dk_z \frac{\sqrt{M_i^{*2}+2\bar \nu |q_i|B}-s\kappa_i B}
{\epsilon^{i}_{k,\bar \nu,s}\sqrt{M_i^{*2}+2\bar \nu |q_i|B}}\\
&\quad \times 
\Bigg\{
\frac{1}{1+e^{\beta(\epsilon^{i}_{k,\bar \nu,s}-\nu_i^{*})}}
+
\frac{1}{1+e^{\beta(\epsilon^{i}_{k,\bar \nu,s}+\nu_i^{*})}}
\Bigg\} \\
&\quad+ \sum_{\bar \nu=1}^{\bar \nu_{max}^{(s=-1)}}M_i^{*}
\int_{0}^{\infty} dk_z \frac{\sqrt{M_i^{*2}+2\bar \nu |q_i|B}-s\kappa_i B}
{\epsilon^{i}_{k,\bar \nu,s}\sqrt{M_i^{*2}+2\bar \nu |q_i|B}} \\
&\quad
\times \Bigg\{
\frac{1}{1+e^{\beta(\epsilon^{i}_{k,\bar \nu,s}-\nu_i^{*})}}
+
\frac{1}{1+e^{\beta(\epsilon^{i}_{k,\bar \nu,s}+\nu_i^{*})}}
\Bigg\}\Bigg],
\end{aligned}
\end{equation}
while for neutral baryons it takes the form
\begin{equation}
\begin{aligned}
\rho_{s,\mathrm{med}}^{i}
= 
M_i^{*}
\sum_{s=\pm1}
\int \frac{d^3k}{(2\pi)^3}
\frac{\sqrt{k_x^2+k_y^2+M_i^{*2}}-s\kappa_i B}
{\epsilon^{i}_{k,s}\sqrt{k_x^2+k_y^2+M_i^{*2}}}\\
&\quad\hspace{-7.67cm} \times \left[
\frac{1}{1+e^{\beta(\epsilon^{i}_{k,s}-\nu_i^{*})}}
+
\frac{1}{1+e^{\beta(\epsilon^{i}_{k,s}+\nu_i^{*})}}
\right].
\end{aligned}
\end{equation}
In this work, the magnetized Dirac sea contribution is included within the weak magnetic field approximation. The baryon propagator is expanded in powers of the magnetic field via $q_iB$ and $\kappa_iB$, retaining terms up to second order. The self-energy contribution, which diverges without a magnetic field, is disregarded due to its negligible size. The finite magnetic field dependent contribution modifies the scalar density through vacuum polarization effects induced by the external magnetic field~\cite{Constantinescu:1972ie}. Consequently, the scalar density receives an additional contribution from the magnetized Dirac sea, which is given by~\cite{Mukherjee:2018ebw,Parui:2022hepph,De:2022hepph,Parui:2022openbottom,Mishra2024Charmonium},

\begin{equation}
\rho_s^{DS,i}
= -\frac{1}{4\pi^2}
\left[
\frac{(q_i B)^2}{3M_i^{*}}
+ \Big\{(\kappa_i B)^2 M_i^{*}
+ |q_i B|(\kappa_i B)\Big\}
\left(
\frac{1}{2}
+ 2\ln\left(\!\frac{M_i^{*}}{M_i}\right)
\right)
\right].
\end{equation}

\section{Magnetic Moment of Baryons}
\label{sec:magnetic moment}
In this section, the magnetic moments of baryons are evaluated within a framework that combines the CQMF model with
the chiral constituent quark model \(\chi\)CQM where the masses of baryons are given as an input to calculate their magnetic moment. The \(\chi\)CQM framework incorporates the chiral symmetry
breaking effects through quark-Goldstone boson interactions. The total magnetic moment of a baryon is written
as the sum of contributions from valence quarks, sea quarks, and the orbital
 angular momentum of sea quarks. 
Sea quarks arise due to chiral fluctuations of valence quarks via the
emission of Goldstone bosons (GBs),
\begin{equation}
q^{\uparrow(\downarrow)} \rightarrow \text{GB}^0 + q'^{\downarrow(\uparrow)}
\rightarrow (q\bar{q}') + q'^{\downarrow(\uparrow)},
\end{equation}
where the emitted GB subsequently splits into a quark-antiquark pair, generating the
sea quark content of the baryon~\cite{Cheng:1995,Cheng:1998PRD,Cheng:1997prl,Linde:1998PRD452,Linde:1998PRD452,Song:1997,Song:1998,Dahiya:2001prd,Dahiya:2004IJMPA,Dahiya:2006IJMPA}. The effective interaction between quarks and the GB is described by the Lagrangian~\cite{NSharma2010}. 
\begin{equation}
\mathcal{L}_{\text{int}}
= g_{15}\, \bar{q}\,(\Phi')\,q ,
\end{equation}
where \(g_{15}\) denotes the quark GB coupling strength and \(\Phi'\) represents the
GB field matrix. The field \(\Phi'\) incorporates the pseudoscalar mesons
arising from the spontaneous breaking of chiral symmetry and is expressed as
\begin{equation}
\resizebox{\textwidth}{!}{$
\small
\Phi' =
\begin{pmatrix}
\frac{\pi^{0}}{\sqrt{2}} + \beta\frac{\eta}{\sqrt{6}} + \zeta\frac{\eta'}{4\sqrt{3}} - \gamma\frac{\eta_c}{4}
& \pi^{+} & \alpha K^{+} & \gamma \bar{D}^{0} \\

\pi^{-}
& -\frac{\pi^{0}}{\sqrt{2}} + \beta\frac{\eta}{\sqrt{6}} + \zeta\frac{\eta'}{4\sqrt{3}} - \gamma\frac{\eta_c}{4}
& \alpha K^{0} & \gamma D^{-} \\

\alpha K^{-} & \alpha \bar{K}^{0}
& -2\beta\frac{\eta}{\sqrt{6}} + \zeta\frac{\eta'}{4\sqrt{3}} - \gamma\frac{\eta_c}{4}
& \gamma D_s^{-} \\

\gamma D^{0} & \gamma D^{+} & \gamma D_s^{+}
& -3\zeta\frac{\eta'}{4\sqrt{3}} + 3\gamma\frac{\eta_c}{4}
\end{pmatrix}.
\notag$}
\end{equation}
The parameters \(\alpha\), \(\beta\), \(\zeta\), and \(\gamma\) encode the effects
of SU(3) and SU(4) symmetry breaking associated with the emission of kaons, eta and charmed mesons, respectively.
The contributions of valence,  sea quarks and orbital angular momentum of sea quarks to the total magnetic moment of a baryon are
expressed as
\begin{equation}
\mu_{B}^{*}
=
\mu_{\text{val}}^{*}
+
\mu_{\text{sea}}^{*}
+
\mu_{\text{orbit}}^{*},
\label{eq:mmtotal}
\end{equation}
where
\begin{equation}
\mu_{\text{val}}^{*}
= \sum_{q=u,d,s,c} \Delta q_{\text{val}}\, \mu_q^{*},
\quad
\mu_{\text{sea}}^{*}
= \sum_{q=u,d,s,c} \Delta q_{\text{sea}}\, \mu_q^{*},
\quad \text{and}
\quad\mu_{\text{orbit}}^*
= \sum_{q=u,d,s,c} \Delta q_{\text{val}}\, \mu^{*}(q^{+} \rightarrow q'^{-}).
\end{equation}
Here, \(\Delta q_{\text{val}}\) and  \(\Delta q_{\text{sea}}\) denote the spin polarizations
of the constituent valence and sea quarks, respectively and are given in Table~\ref{tab_magmoments_eqn} in the appendix for each decuplet baryons. The term \(\mu^{*}(q^{+} \rightarrow q'^{-})\) denotes the orbital magnetic moment
associated with a specific chiral fluctuation process. To incorporate quark confinement effects together with relativistic corrections to the
effective quark magnetic moments, the following expressions are employed for the
constituent quarks of a given baryon~\cite{Gridhar:2015prd}
\begin{equation}
\mu_d^{*} = -\left(1 - \frac{\Delta M}{M_i^{*}}\right), \qquad
\mu_s^{*} = -\frac{m_u^{*}}{m_s^{*}}
\left(1 - \frac{\Delta M}{M_i^{*}}\right),
\label{eq:mud}
\end{equation}
\begin{equation}
\mu_u^{*} = -2\,\mu_d^{*}, \hspace{0.4cm}\text{and}   \quad
\mu_c^{*} = -\frac{2m_u^{*}}{m_c^{}}\,\mu_d^{*}.
\label{eq:muu}
\end{equation}
Here, \(M_i^{*}\) is evaluated using
Eq. (\ref{baryonmass}), and \(\Delta M = M_{\text{vac}} - M_i^{*}\), where \(M_{\text{vac}}\) represents
the vacuum mass of the baryon.
In the present calculations, the orbital contributions from
the four quark flavors \(u\), \(d\), \(s\), and \(c\) are included, and the explicit
expressions for the transition magnetic moments are given in Ref.~\cite{NSharma2010}.


\section{Results and discussion} 
\label{sec:results}
In this section, we present the results of the medium modified masses and magnetic moments of decuplet baryons in magnetized strange matter with finite isospin asymmetry $I_a$, strangeness fraction $f_s$ and baryonic densities $\rho_B$ at temperature $T$. In Eqs.  (\ref{eq:mud}) and   (\ref{eq:muu}), the value of the magnetic 
moments of constituent quarks are determined from the effective
masses of both quarks and baryons, which themselves are the functions of the scalar fields. Consequently, analyzing the magnetic field dependence of scalar fields and baryon masses are crucial for evaluating the medium modification of magnetic moments of baryons.
\begin{table}[h]
\centering
\renewcommand{\arraystretch}{1.9}
\setlength{\tabcolsep}{13pt}
\begin{tabular}{|c|c|c|c|c|c|}
\hline
$k_0$ & 4.94 & $k_1$ & 2.12 & $k_2$ & $-10.16$ \\
\hline
$k_3$ & $-5.38$ & $k_4$ & $-0.06$ & $g_4$ & 37.5 \\
\hline
$d$ & 0.182 & $\rho_0$ (fm$^{-3}$) & 0.16 & $m_v$ (MeV) & 673.648 \\
\hline
$f_\pi$ (MeV) & $-93$ & $f_K$ (MeV) & 115 & $m_\pi$ (MeV) & 139 \\
\hline
$m_K$ (MeV) & 496 & $m_\omega$ (MeV) & 783 & $m_\phi$ (MeV) & 1020 \\
\hline
$\sigma_0$ (MeV) &$ -93$ & $\zeta_0$ (MeV) & $-96.87$ & $\chi_0$ (MeV) & 254.38 \\
\hline
$g_\sigma^{u,d}$ & 2.72 & $g_\sigma^{s}$ & 0 & $g_\zeta^{u,d}$ & 0 \\
\hline
$g_\zeta^{s}$ & 3.847 & $g_\delta^{u,d}$ & 2.72 & $g_\delta^{s}$ & 0 \\
\hline
$g_\omega^{u,d}$ & 3.23 & $g_\omega^{s}$ & 0 & $g_\rho^{u,d}$ & 3.23 \\
\hline
$g_\rho^{s}$ & 0 &$g_\phi^{u,d}$ & 0 & $g_\phi^{s}$ & 4.57\\
\hline
$a$ & 0.12 & $a\alpha^2$ = $a\beta^2$ & 0.0243 & $a\zeta^2$ & 0.0053 \\
\hline
$a\gamma^2$ & 0.0014 & $M_\eta$ (MeV) & 547.862 & $M_{\eta'}$ (MeV) & 957.78 \\
\hline
$M_{\eta_c}$ (MeV) & 2983.5 & $M_D$ (MeV) & 1869 & $M_{D_s}$ (MeV) & 1968 \\
\hline
 $m_c$ (MeV) & 1270 & - & - & - & - \\
\hline
\end{tabular}
\caption{Various parameters and coupling constants utilized in the present calculations~\cite{pwang2003}.}
\label{params}
\end{table}

\begin{table}[h]
\centering
\renewcommand{\arraystretch}{1.4}
\setlength{\tabcolsep}{45pt}
\begin{tabular}{|c|c|c|c|}
\hline
$g_{\sigma p}$ & 6.64 & $g_{\omega p}$ & 9.69 \\\hline
$g_{\zeta p}$ & 0 & $g_{\rho p}$ & 8.89 \\\hline
$g_{\delta p}$ & 2.72 & $g_{\phi p}$ & 0 \\
\hline
$g_{\sigma n}$ & 6.64 & $g_{\omega n}$ & 9.69 \\\hline
$g_{\zeta n}$ & 0 & $g_{\rho n}$ & 8.89 \\\hline
$g_{\delta n}$ & 2.72 & $g_{\phi n}$ & 0 \\
\hline
$g_{\sigma \Lambda}$ & 5.44 & $g_{\omega \Lambda}$ & 8.85 \\\hline
$g_{\zeta \Lambda}$ & 3.85 & $g_{\rho \Lambda}$ & 0 \\\hline
$g_{\delta \Lambda}$ & 0 & $g_{\phi \Lambda}$ &$ -4.57$ \\
\hline
$g_{\sigma \Sigma}$ & 5.44 & $g_{\omega \Sigma}$ & 12.28 \\\hline
$g_{\zeta \Sigma}$ & 3.85 & $g_{\rho \Sigma}$ & 12.28 \\\hline
$g_{\delta \Sigma}$ & 2.72 & $g_{\phi \Sigma}$ & $-4.57$ \\
\hline
$g_{\sigma \Xi}$ & 2.72 & $g_{\omega \Xi}$ & 7.27 \\\hline
$g_{\zeta \Xi}$ & 7.69 & $g_{\rho \Xi}$ & 3.23 \\\hline
$g_{\delta \Xi}$ & 2.72 & $g_{\phi \Xi}$ &$ -9.14$ \\
\hline
\end{tabular}
\caption{Baryon-meson coupling constants used in the present calculation.}
\label{tab:baryoncouplings}
\end{table}

\subsection{In-medium scalar fields in magnetized strange matter}
\label{sec_fields}

\begin{figure}
    \centering
    \includegraphics[width=0.80\linewidth]{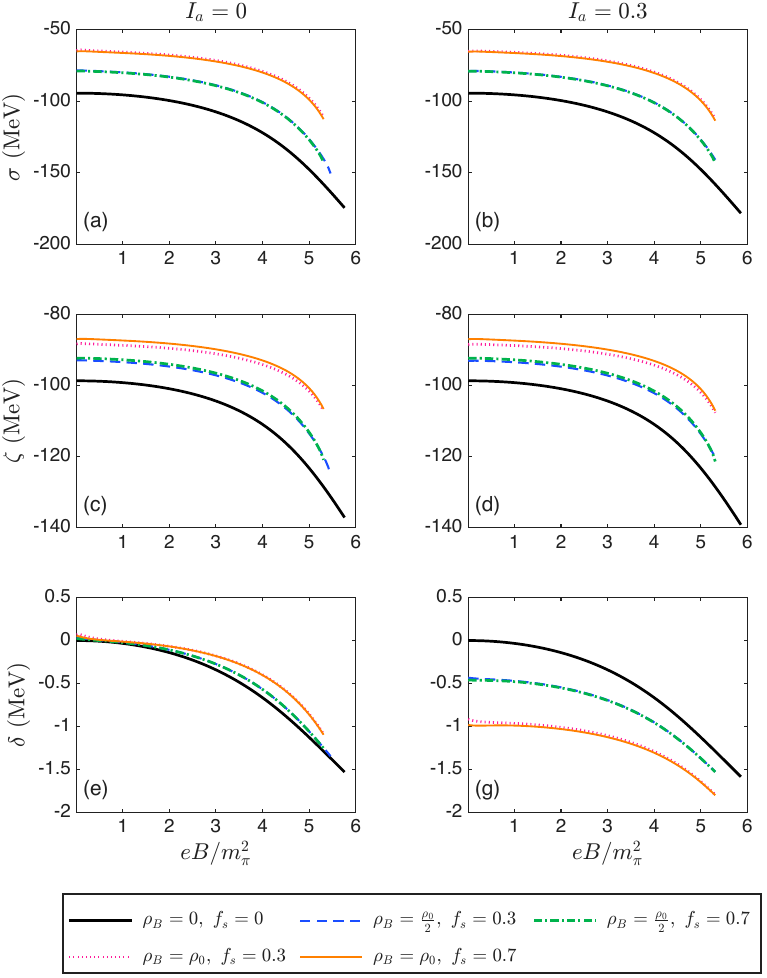}
    \caption{Variation of scalar fields ($\sigma,\zeta$ and $\delta$) with magnetic field $eB/m_\pi^2$ for isospin symmetric $I_a=0$ [Subplots (a), (c), and (e)] and isospin asymmetric $I_a=0.3$ [subplots (b), (d), and (f)] at temperature T=100 MeV. Each subplot shows the results corresponding to different baryonic density $\rho_B$, strangeness fraction $f_s$.}
    \label{fields}
\end{figure}

Figure~\ref{fields} shows the variation of the scalar fields $\sigma$, $\zeta$, and $\delta$ as functions of the magnetic field strength $eB/m_\pi^2$ for different baryon densities ($\rho_B=\frac{\rho_0}{2},\rho_0$) and strangeness fractions ($f_s=0,0.3,0.7$) at temperature $T=100$ MeV. The left and right panels correspond to isospin-symmetric matter ($I_a=0$) and isospin-asymmetric matter ($I_a=0.3$), respectively.
For given isospin asymmetry $I_a$, all the fields become progressively more negative across all medium parameters with increase in magnetic field. From Figs.~\ref{fields} (a) and (b), we observe that the magnitude of the non-strange scalar field $\sigma$ ($\sim m_u \langle u\bar{u}\rangle$ + $m_d\langle d\bar{d}\rangle $) increases with increase in magnetic field within the strange medium. This finite increase as demonstrated in studies in Ref.~\cite{Mishra2024Charmonium} results from the contribution of the Dirac sea, which shows negligible variation when the contribution of Dirac sea is not considered. The solutions of the scalar fields exist only up to a critical magnetic field $eB \sim 5.5m_\p^2$ in the strange matter at finite densities.
In Fig.~\ref{fields} (a) for isospin symmetric case ($I_a=0$), at zero baryonic density ($\rho_B = 0$), 
the magnitude of the $\sigma$ field remains relatively larger, with a value $-94.24$ MeV, compared to it's value at finite baryonic density $\rho_B = \frac{\rho_0}{2}$ ($\rho_0$), 
where it changes to $ -64.44$ ($-78.39$) MeV at $f_s=0.3$ and zero magnetic field.
The relative increase in magnitude of the field as a function of magnetic field indicates the magnetic catalysis behavior and the rate of change in magnitude of $\sigma$ field with magnetic field is more noticeable beyond $eB=4m_\p^2$. The inclusion of a finite strangeness fraction $f_s$ has a negligible impact on the behavior of the $\sigma$ field, indicating its weak sensitivity to the strangeness content of the medium.
 As moving from isospin symmetric to asymmetric $I_a=0.3$ matter, we observe a small shift of 0.27 MeV at $\rho_B = \rho_0$, $f_s=0.7$ and zero magnetic field in the magnitude of $\sigma$ field. 
 
The strange scalar field $\zeta$ ($\sim m_s \langle s\bar{s} \rangle$) also shows an increase the magnitude with magnetic field which lies in a narrower range compared to $\sigma$ field as shown in the Figs.~\ref{fields} (c) and (d). For example, in Figs.~\ref{fields} (a) and (c) for baryonic density $\rho_B=\frac{\rho_0}{2}$, $I_a=0$, and $f_s=0.3$ the magnitude of scalar field $\zeta(\sigma)$ is observed to be $-92.35$ ($-78.86$) MeV at zero magnetic field, $-96.59$ ($-89.11$) MeV at $eB=3m_\p^2$ and $-113.35$ ($-127.22$) MeV at $eB=5m_\p^2$. We observe a small decrease of $0.83\%$ in magnitude of $\zeta$ field as $f_s$ is increased from 0.3 to 0.7 at $\rho_B=\rho_0$ and $eB=5m_\pi^2$, indicating the response of the strange condensate in the medium. As observed for the $\sigma$ field, the magnitude of the $\zeta$ field decreases with increasing baryonic density. For $f_s=0.7$, the corresponding values of the $\zeta$ field at $\rho_B=\frac{\rho_0}{2}$ and $\rho_0$ are $-92.35$ ($-113.35$) MeV and $-86.98$ ($-100.89$) MeV at $eB=0m_\pi^2$ ($5m_\pi^2$), respectively. With increase in $I_a$ from 0 to 0.3, a negligible rise in magnitude of value 0.007 MeV is observed, reflecting the $\zeta$ field's weak sensitivity to isospin asymmetry.  
 
 The Figs.~\ref{fields} (e) and (f) shows the variation of scalar-isovector field $\delta$ as a function of magnetic field. For isospin symmetric matter $I_a = 0$, the $\delta$ field is observed to be nearly zero for the vacuum ($\rho_B = 0$) and finite density ($\rho_B =\frac{\rho_0}{2},\rho_0$) cases across all strangeness fractions $f_s$ at zero magnetic field as expected. This is because, in the isospin symmetric matter, the number of $u$ and $d$ quarks are equal therefore, the $\delta$ field vanishes. However, as the magnetic field strength is increased, the magnitude of the $\delta$ field also increases, showing nearly identical behavior across different baryonic densities and strangeness fractions. The Fig.~\ref{fields} (f) shows a pronounced difference at finite isospin asymmetry ($I_a=0.3$) in the magnitude of the $\delta$ field comparing to other fields. Unlike $\sigma$ and $\zeta$ fields, the magnitude of $\delta$ field increases with baryon density,
being largest at $\rho_B = \rho_0$ and it reflects
the enhanced isospin splitting effects.
The $\delta$ field is also moderately sensitive to strangeness fraction. At $\frac{\rho_0}{2}$ and $I_a=0.3$, the $\delta$ field changes from $-0.957$ MeV to $-0.960$ MeV as $f_s$ is increased from 0.3 to 0.7.
 Thus, the magnetic response of the scalar fields are density dominated,
 with strangeness and isospin asymmetry acting as a secondary factor.

\begin{figure}
    \centering
    \includegraphics[width=0.77\linewidth]{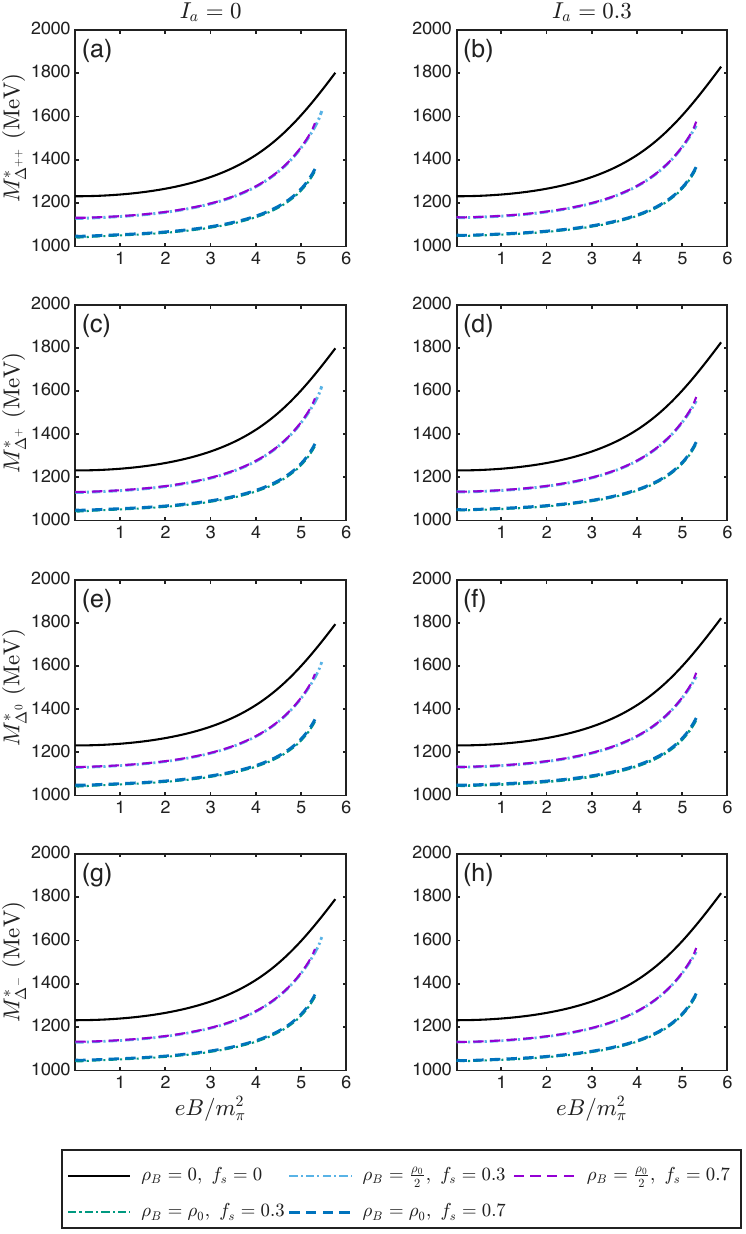}
    \caption{Variation of masses of $\Delta$ baryons with magnetic field $eB/m_\pi^2$ for isospin symmetric $I_a=0$ [Subplots (a), (c), (e) and (g)] and isospin asymmetric $I_a=0.3$ [subplots (b), (d), (f) and (h)] at T=100 MeV. Each subplots shows the result corresponding to different baryonic density $\rho_B$, strangeness fraction $f_s$.}
    \label{fig_mass_delta}
\end{figure}
\subsection{In-medium masses of decuplet baryons}
\label{masses_results}

\begin{figure}
    \centering
    \includegraphics[width=0.85\linewidth]{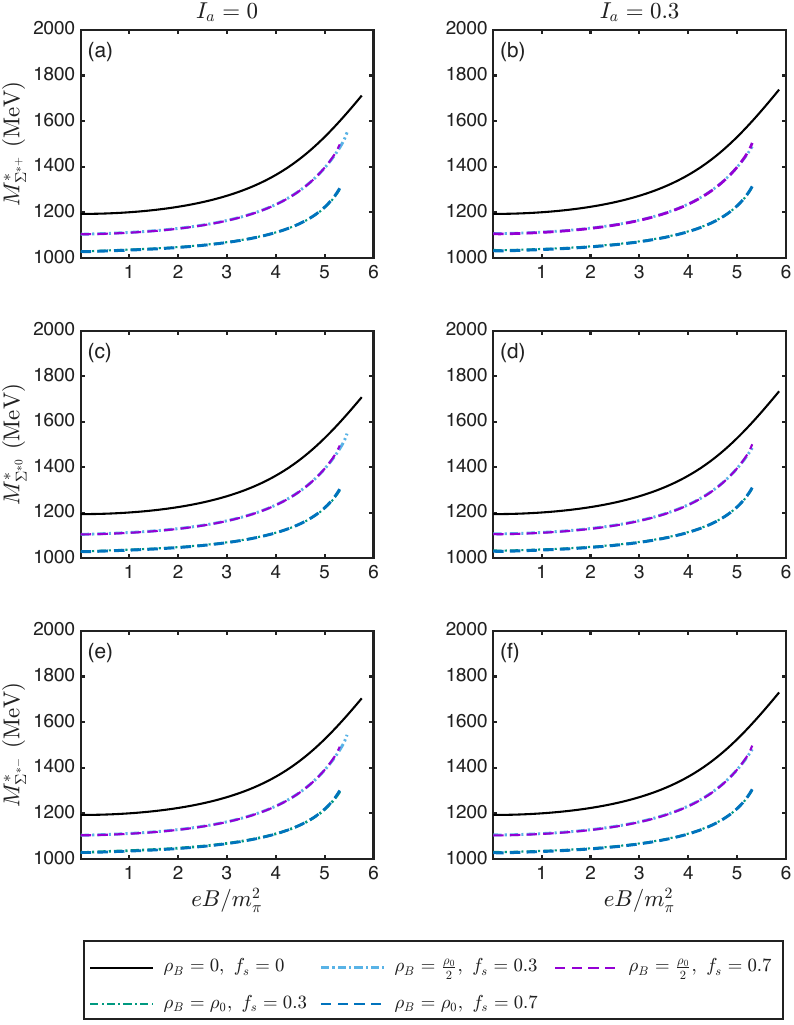}
    \caption{Same as Fig.~\ref{fig_mass_delta}, for $\Sigma^*$ baryons}
    \label{fig_sigma_mass}
\end{figure}

\begin{figure}
    \centering
    \includegraphics[width=0.85\linewidth]{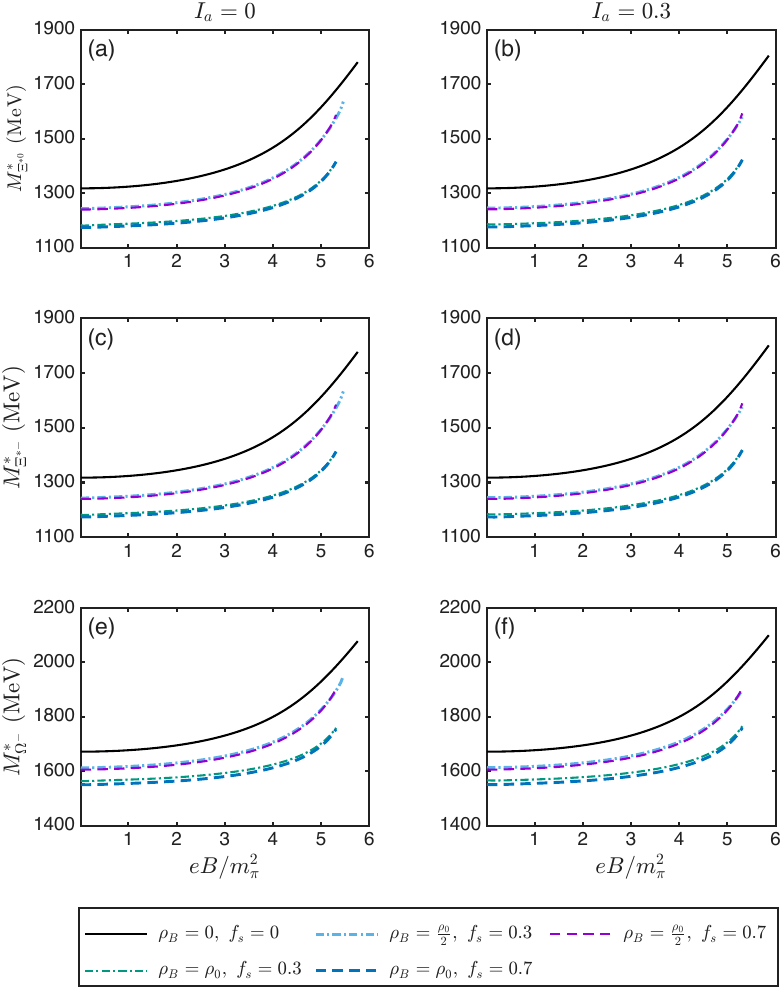}
    \caption{Same as Fig~.~\ref{fig_mass_delta}, for $\Xi^{*}$ baryons [subplots (a), (b), (c) and (d)] and $\Omega^-$ baryons [subplots (e) and (f)].}
    \label{fig_mass_xi}
\end{figure}

Now we discuss the behavior of the effective masses of the decuplet baryons as a function of  magnetic field $eB/m_\p^2$ for different baryonic densities ($\rho_B=\frac{\rho_0}{2},\rho_0$) and strangeness fraction ($f_s= 0.3 , 0.7)$ at finite isospin asymmetry $I_a = 0$ (left panel) and $I_a = 0.3 $ (right panel) at temperature $T=100$ MeV. The effective masses of the decuplet baryons are obtained from Eq.(~\ref{baryonmass}), where the values of $E_{i,spin}$ in Eq.(~\ref{energyeqn}) is fitted to get the vacuum masses of the respective baryons.  In Figs.~\ref{fig_mass_delta}, ~\ref{fig_sigma_mass} and ~\ref{fig_mass_xi}, we have shown the variation of in-medium masses of $\Delta^{++,+,0,-}$, $\Sigma^{*\pm,0}$, $\Xi^{*-,0}$ and $\Omega^-$ baryons as a function of magnetic field. For a quantitative comparison, the corresponding masses of the decuplet baryons at different magnetic fields $eB/m_\pi^2$, baryon densities $\rho_B$, with $f_s=0.3$, $I_a=0.3$, and $T=100$ MeV are listed in Table~\ref{tab:baryon_masses}.

From Fig.~\ref{fig_mass_delta}, we can see the variation of effective masses of 
$\Delta$ baryons in magnetized strange matter.
Since these baryons consist purely of light quarks ($u$ and $d$ quarks),
their masses are primarily governed by the $\sigma$ field,
with additional splitting arising from the $\delta$ field in isospin asymmetric matter.
For isospin symmetric matter ($I_a = 0$),
the $\delta$ field vanishes almost identically, therefore the masses of all four $\Delta$ states remain almost degenerate
at a given density and magnetic field as depicted in Fig~\ref{fig_mass_delta}.
The mass modifications are therefore entirely controlled
by the $\sigma$ field and magnetic effects,
such as Landau quantization and anomalous magnetic moments. From the Fig.~\ref{fig_mass_delta} and Table~\ref{tab:baryon_masses}, we observe that the magnitude of the masses of $\Delta$ baryons are increasing with magnetic field reflecting magnetic catalysis,
where the magnitude of the $\sigma$ field increases,
leading to slightly enhanced constituent quark masses.
For $I_a=0$, at $f_s=0.7$ and finite densities, $\rho_B = \frac{\rho_0}{2}$ and $\rho_0$,
the $\Delta^{++}$ baryon mass decreases by $\sim8\%$ and $\sim15\%$ respectively from their value at $\rho_B=0$ and $f_s=0$.
The influence of strangeness fraction $f_s$
is comparatively weaker for the $\Delta$ baryons,
since they do not contain strange quarks. Comparing to the isospin asymmetric medium in Fig.~\ref{fig_mass_delta} (right panel), the trend remains the same but we observe a small increase in magnitude of all $\D$ baryon masses.

The $\Sigma^*$ baryons contain one strange quark and two light quarks.
Consequently, their effective masses depend on both
the $\sigma$
and $\zeta$ fields. From Fig.~\ref{fig_sigma_mass},
the effective masses of $\Sigma^*$ baryons exhibit an increase
with magnetic field showing similar trend as observed for $\Delta$ baryons.
At finite densities $\rho_B = \frac{\rho_0}{2}$ and $\rho_0$ and, strangeness fraction $f_s=0.3$,
the mass of $\Sigma^*$ baryons decreases by $\sim7\%$ and $\sim14\%$ from vacuum value of 1193.25 MeV at zero magnetic field for $I_a=0$. As we move from $I_a=0$ to $0.3$, we observe a mild change of $0.88$ MeV in mass at $\frac{\rho_0}{2}$ and $f_s =0.3$ as shown in the Fig.~\ref{fig_sigma_mass} since only two of the quarks are sensitive to isospin asymmetry. The effect of strangeness fraction is also very weak, as there is only one quark in the $\Sigma^*$ which is sensitive to strangeness fraction.. 

The effective mass of $\Xi^*$ baryons which contains two strange quarks and one light quark increases with increase in magnetic field as depicted in Figs.~\ref{fig_mass_xi} (a), (b), (c) and (d). At zero density the mass is $1317$ MeV and at finite densities,
a decrease in mass is observed for both values of the strangeness fraction, around $\sim5\%$ and $\sim 10\%$ for $\rho_0$ and $\frac{\rho_0}{2}$ respectively at zero magnetic field.
The influence of strangeness fraction $f_s$
is also slightly higher compared to $\Delta$ and $\Sigma^*$ baryons,
since the strange condensate ($\zeta$ field) directly controls
two-thirds of the baryon mass contribution. Therefore, the effect of isospin asymmetry is also very small for $\Xi^*$ baryons. 

As observed in Figs.~\ref{fig_mass_xi} (e) and (f), the mass of $\Omega^-$ baryon also increases with increase in magnetic field as discussed for other decuplet baryons. In Fig.~\ref{fig_mass_xi} (e), for isospin symmetric case, the mass of $\Omega^-$ baryon takes the value 1672.02 MeV at $\rho_B=f_s=0$ and as we increase the density to $\frac{\rho_0}{2}$ and $\rho_0$, the mass decreases by 58.03 and 125.07 MeV respectively at $f_s=0.3$ and zero magnetic field as given in Table~\ref{tab:baryon_masses}. The $\Omega^-$ baryon does not experience
isospin-induced splitting as it does not contain any light quarks,
and its magnetic response is governed primarily by
Landau quantization and anomalous magnetic moment effects. As we increase $f_s$ from 0.3 to 0.7, the mass of $\Omega^-$ baryon drops by $5.9$ MeV at $\rho_B = \rho_0$ and $I_a=0.3$. The $\Omega^-$ baryon consists entirely of strange quarks and therefore, the effect of strangeness fraction is more pronounced compared to other decuplet baryons.

\begin{table}[ht]
\centering
\renewcommand{\arraystretch}{1.8}
\setlength{\tabcolsep}{15pt}
\begin{tabular}{|c|c|c|c|c|}
\hline
& \multicolumn{2}{c|}{$\rho_B=\rho_0/2$ (MeV)} 
& \multicolumn{2}{c|}{$\rho_B=\rho_0$ (MeV)} \\

\cline{2-5}

Baryon 
& $eB=0$ $m_\pi^2$ 
& $eB=5$ $m_\pi^2$ 
& $eB=0$ $m_\pi^2$ 
& $eB=5$ $m_\pi^2$ \\
\hline

$\Delta^{++}$ & 1134.28 & 1461.06 & 1050.54 & 1269.20 \\
\hline
$\Delta^{+}$  & 1133.37 & 1457.85 & 1048.72 & 1265.55 \\\hline

$\Delta^{-}$  & 1132.46 & 1454.63 & 1046.89 & 1261.89 \\\hline

$\Delta^{0}$  & 1131.55 & 1451.42 & 1045.08 & 1258.24 \\\hline

$\Sigma^{*+}$ & 1107.89 & 1400.37 & 1035.61 & 1230.20 \\\hline
$\Sigma^{*-}$ & 1106.06 & 1393.91 & 1031.95 & 1222.85 \\\hline
$\Sigma^{*0}$ & 1106.98 & 1397.14 & 1033.78 & 1226.53 \\\hline

$\Xi^{*-}$    & 1244.87 & 1499.74 & 1164.166 & 1350.38 \\\hline
$\Xi^{*0}$    & 1245.78 & 1500.97 & 1185.99 & 1354.03 \\\hline

$\Omega^{-}$  & 1613.99 & 1830.47 & 1546.95 & 1707.53 \\\hline

\end{tabular}
\caption{Masses of decuplet baryons (in MeV) at different magnetic fields $eB/m_\p^2$, baryon densities $\rho_B$, strangeness fraction $f_s= 0.3$, isospin asymmetry $I_a=0.3$ and temperature $T=100$ MeV.}
\label{tab:baryon_masses}
\end{table}

The existing studies on decuplet baryon masses are found to decrease monotonically with increasing baryonic density consistent with previous CQMF studies in symmetric nuclear and hyperonic matter~\cite{Singh:2020epjp,Singh:2019epjp,Singh:2018epja}.
The role of isospin asymmetry in the medium has been studied using the CQMF model 
in Refs.~\cite{Kumar:2023owb, Singh:2017cpc}, where the scalar-isovector field 
$\delta$ is found to induce charge state splitting among the multiplet members, 
with the splitting being most pronounced for baryons composed entirely of light 
quarks. 
The 
inclusion of Dirac sea contributions are essential for magnetic 
catalysis of scalar condensates in magnetized 
matter~\cite{Menezes:2009prc2, Mukherjee:2018prd, Aguirre:2016epja, Aguirre:2019prc} and at the 
quark level, the enhancement of the scalar condensate $\sigma$ with increasing 
magnetic field is due to the phenomenon of magnetic catalysis of chiral symmetry 
breaking established in Refs.~\cite{Gusynin:1996npb, Miransky_2002}, where a constant 
magnetic field acts as a catalyst of dynamical chiral symmetry breaking. These existing results are consistent with the present findings.

\subsection{In-medium magnetic moments of decuplet baryons}

\begin{figure}
    \centering
    \includegraphics[width=0.85\linewidth]{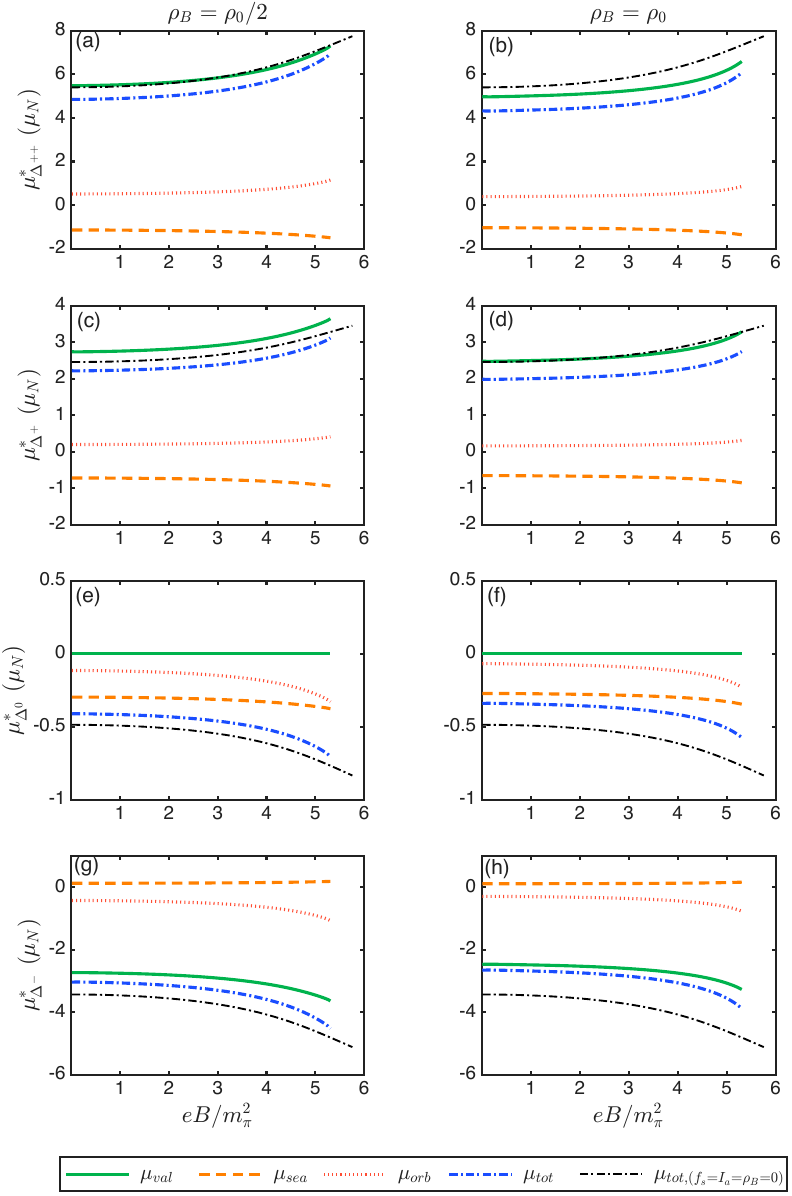}
    \caption{In-medium magnetic moment variation of $\Delta$ baryons as a function of magnetic field $eB/m_\pi^2$ for density $\rho_B = \frac{\rho_0}{2}$ [Subplots (a), (c), (e) and (g)] and $\rho_B = \rho_0$[subplots (b), (d), (f) and (h)] at $f_s=0.7$, $I_a=0.3$ and T=100 MeV. The black dash-dotted curve corresponds to the total magnetic moment calculated for $\rho_B=0$, $f_s=0$ and $I_a=0$. Each subplots shows the result corresponding to different contributions to magnetic moments.}
\label{fig_mm_delta}
\end{figure}

\begin{figure}
    \centering
    \includegraphics[width=0.87\linewidth]{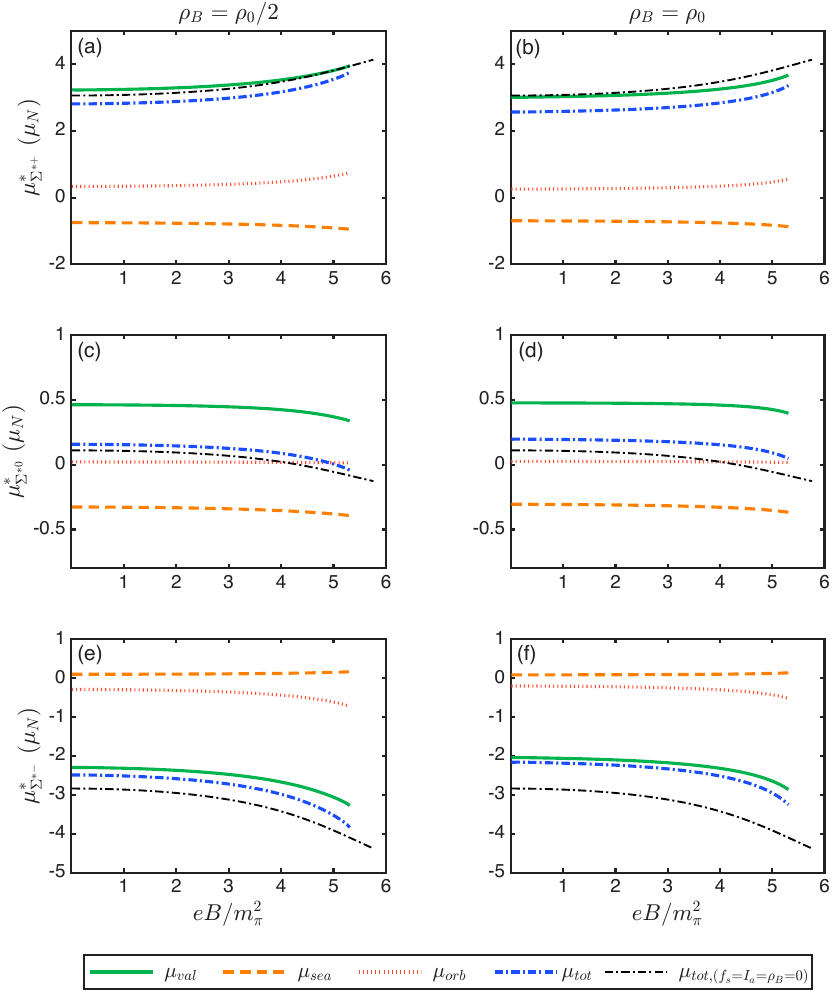}
    \caption{In-medium magnetic moment variation of $\Sigma^{*}$ baryons as a function of magnetic field $eB/m_\pi^2$ for density $\rho_B = \frac{\rho_0}{2}$ [Subplots (a), (c) and (e)] and $\rho_B = \rho_0$[subplots (b), (d) and (f)] at $f_s=0.7$, $I_a=0.3$ and T=100 MeV. The black dash-dotted curve corresponds to the total magnetic moment calculated for $\rho_B=0$, $f_s=0$ and $I_a=0$. Each subplots shows the result corresponding to different contributions to magnetic moments.}
\label{fig_mm_sigma}
\end{figure}

\begin{figure}
    \centering
    \includegraphics[width=0.87\linewidth]{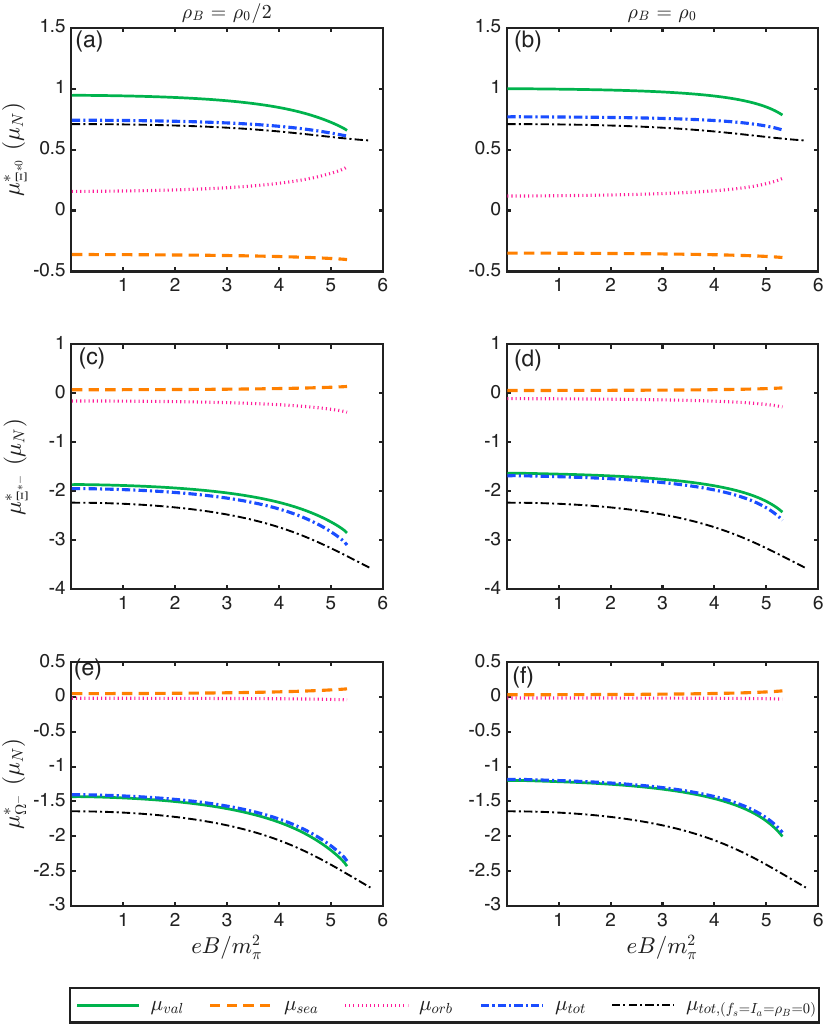}
    \caption{In-medium magnetic moment variation of $\Xi^{*}$ and $\Omega^-$ baryons as a function of magnetic field $eB/m_\pi^2$ for density $\rho_B = \frac{\rho_0}{2}$ [Subplots (a), (c), (e)] and $\rho_B =\rho_0$[subplots (b), (d), (f)] at $f_s=0.7$, $I_a=0.3$ and T=100 MeV. The black dash-dotted curve corresponds to the total magnetic moment calculated for $\rho_B=0$, $f_s=0$ and $I_a=0$. The different contributions to magnetic moments are shown separately in each subplots.}
\label{fig_mm_xi}
\end{figure}

In this section we present the results of magnetic moments of decuplet baryons and their behaviour in magnetized strange matter. Figures~\ref{fig_mm_delta}, ~\ref{fig_mm_sigma} and~\ref{fig_mm_xi}  show the total effective magnetic
moments and the individual contributions of valence, sea and orbital angular momentum of sea quarks of decuplet baryons for different baryonic densities $\rho_B = \frac{\rho_0}{2}$ (left panel) and $\rho_B = \rho_0$ (right panel) at strangeness fraction $f_s=0.7$ and isospin asymmetry $I_a=0.3$.
The underlying phenomena of the magnetic moment enhancement at finite density is due to the medium suppressing the effective quark masses $m^*_q$, thereby effective baryon masses and the constituent baryon magnetic moments as a function of density. Whereas the magnetic moment tends to increase in magnitude with magnetic field due to magnetic catalysis effects. The tables~\ref{tab_magmoments1}, ~\ref{tab_magmoments2} and ~\ref{tab_magmoments3} shows the values of magnetic moments of all decuplet baryons at different baryonic densities, strangeness fraction and isospin asymmetry for different values of magnetic field. 

The Fig.~\ref{fig_mm_delta} shows the variations of magnetic moment of $\Delta$ baryons with magnetic field. The $\Delta$ baryons show the most pronounced magnetic moment variation with  magnetic field, as expected from the strong medium sensitivity of the $\sigma$ field. The total magnetic moment of $\Delta^{++,\pm,0,}$ baryons are observed to increase with increasing magnetic field. From Fig.~\ref{fig_mm_delta} (a), at $\rho_B=\frac{\rho_0}{2}$, the valence contribution is dominant and positive ($5.488\mu_N$), primarily driving the increase in the baryon magnetic moment, while the sea contribution is negative ($-1.143\mu_N$) and the orbital contribution is positive but comparatively small ($0.511\mu_N$) at zero magnetic field. Due to the effect of these three contributions the net magnetic moment turns out to be $4.856$$\mu_N$ at zero magnetic field whereas the value of magnetic moment is 5.408$\mu_N$ at vacuum, i.e., $\rho_B=0,f_s=0$ and $I_a=0$.  As in the Fig.~\ref{fig_mm_delta} (b), when the density is increased to $\rho_0$, the total magnetic moment of $\Delta^{++}$ baryon decreases to 4.328 $\mu_N$ at zero magnetic field. In Figs.~\ref{fig_mm_delta} (c) and (d), at zero magnetic field, the $\Delta^+$ baryon has a positive magnetic moment value $2.220$$\mu_N$ at $\rho_B = \frac{\rho_0}{2}$ and decreases by $0.23\mu_N$ when the density increases to $\rho_B=\rho_0$. The magnetic moment variation with density and magnetic fields are qualitatively similar to $\Delta^{++}$ baryon. As shown in Figs.~\ref{fig_mm_delta}(e) and (f), the valence quark contribution to the magnetic moment of the $\Delta^{0}$ baryon is zero. This results from the cancellation between the positive contribution of the single $u$ quark and the combined negative contributions of the two $d$ quarks.Therefore, the total magnetic moment is primarily controlled by sea quarks and hence the total magnetic moment follows an opposite trend as compared to other $\D^{++}$ and $\D^{+}$ with magnetic field. The $\Delta^{-}$ baryon also follows the same trend as $\Delta^0$ baryon as shown in the Figs.~\ref{fig_mm_delta} (g) and (h). The $\Delta^{-}$ baryon has a negative magnetic moment and it's magnitude tends to increase with increase in magnetic field and decrease as we increase the density. The sea contribution is small and positive but the negative valence contribution drives the total magnetic moment and takes the value $-3.434$$\mu_N$ at $\rho_B=I_a=f_s=0$ and zero magnetic field.


For $\Sigma^{*+}$ baryons $(uus)$, as shown in 
Figs.~\ref{fig_mm_sigma} (a) and (b), the two $u$ quarks dominate therefore, 
valence contribution is large and positive compared to other contribution to the total magnetic moment. Therefore, the
resulting total magnetic moment of the $\Sigma^{*+}$ baryon is positive and increases with increasing magnetic field. Upon increasing the density to $\rho_0$, the magnitude of 
total magnetic moment decreases by a value 0.241$\mu_N$. 
For $\Sigma^{*0}$ $(uds)$, the magnetic moment decreases with increase in magnetic field and as we increase the density from $\rho_B=\frac{\rho_0}{2}$ to $\rho_0$ the value of the magnetic moment increases by 0.039$\mu_N$ which follows an opposite trend compared to $\Sigma^{*+}$ baryons. Since the
contributions from the $u$, $d$ and $s$ quarks partially cancel and therefore, the total magnetic moment approaches zero at large magnetic field as depicted in Figs.~\ref{fig_mm_sigma} (c) and (d). For $\Sigma^{*+}$ and $\Sigma^{*0}$ baryons the sea contribution is negative, but 
for $\Sigma^{*-}$ baryon $(dds)$, the sea contribution is small but positive as shown in Figs.~\ref{fig_mm_sigma} (e) and (f). For $\Sigma^{*-}$ baryons the two 
$d$ quarks and $s$ quark contribute negatively to the total magnetic moment. Consequently, the total magnetic moment is found to be $-2.491$ ($-2.839$)$\mu_N$ at $\rho_B=\frac{\rho_0}{2}$ ($\rho_B=0$) and decreases to $-2.165$$\mu_N$ as we increase the baryon density to $\rho_B=\rho_0$ in the absence of magnetic field. 

The $\Xi^*$ baryons carry two strange quarks and one light quark. Figures~\ref{fig_mm_xi} (a) and (b), indicates that the total magnetic moment of $\Xi^{*0}$ baryon exhibit a very weak sensitivity to the magnetic field. At baryonic 
density $\frac{\rho_0}{2}$, the sea quark contribution to the total magnetic moment is $-0.361$$\mu_N$ where it decreases by 0.012$\mu_N$ as density is increased to $\rho_0$, whereas all other magnetic moment contributions give positive contribution to the total magnetic moment. The density dependence of magnetic moment of $\Xi^{*0}$ baryons is weak of order $\sim 0.027\mu_N$ as we increase density from $\rho_B=\frac{\rho_0}{2}$ to $\rho_0$, compared to $\D$ and $\Sigma^*$ baryons at zero magnetic field. Moreover, for $\Xi^{*-}$ baryons, as shown in Figs.~\ref{fig_mm_xi} (c) and (d), the single $d$ quark contributes a negative valence moment, and the two $s$ 
quarks reinforce this negative sign. The value of total magnetic moment is therefore 
$-1.947$ ($-1.685$)$\mu_N$ at $\rho_B = \frac{\rho_0}{2}$ ($\rho_0$) at zero magnetic field and their values increases as we increase the magnetic field. The valence term and total magnetic moment
curves nearly coincident since both the sea and orbital contributions are small.

From Figs.~\ref{fig_mm_xi} (e) and (f), we observe that sea and orbital contributions are both close to zero as the $\Omega^-$ baryon has $sss$ quark content and the dominant sea polarization come from light quarks. The magnitude of $\mu^{*}_{\Omega^{-}}$ increases monotonically with 
magnetic field. At vacuum, the value of total magnetic moment of $\Omega^-$ baryon is$ -1.642$$\mu_N$ where it changes to -1.405 and $-1.187$$\mu_N$ as we increase the density to $\rho_B=\frac{\rho_0}{2}$ and $\rho_0$, respectively. 

Within the CQMF framework, the magnetic moments of octet and decuplet baryons have been investigated in Refs.~\cite{Singh:2017cpc,Singh:2019epjp,Singh:2018epja}, demonstrating a strong dependence on the baryon density and isospin asymmetry of the medium. The study was subsequently extended to strange hadronic matter in Ref.\cite{Kumar:2023owb}, where the decuplet baryon magnetic moments were shown to be sensitive to the baryon density, isospin asymmetry, temperature, and strangeness fraction. Using the quark meson coupling (QMC) model~\cite{Tsushima:2022ptep}, the in-medium magnetic moments of octet, decuplet, and low-lying charm and bottom baryons have also been investigated, with the magnetic moments generally found to be enhanced in the medium relative to their free space values. The magnetic moments of decuplet baryons have also been calculated within chiral perturbation theory ($\chi$PT) and lattice QCD (LQCD), where the magnetic moment of the $\Delta^{++}$ baryon was found to be $5.390\mu_N$ and $5.240\mu_N$ in $\chi$PT~\cite{FloresMendieta:2009,Geng:2009} and LQCD~\cite{Leinweber1992} respectively. The 
influence of medium-modified baryonic magnetic moments on neutron star 
properties under strong magnetic fields has been analyzed in 
Ref.~\cite{Ryu:2010prc}, where the AMM of baryons is found to significantly 
alter the equation of state at high densities. All these results are in good agreement with the results obtained in the present study.

\begin{table}[h!]
\centering
\renewcommand{\arraystretch}{2}
\setlength{\tabcolsep}{0.70pt}
\begin{tabular}{|c|c|c|c|c|c|c|c|c|c|c|}
\hline 
 & Exp. values~\cite{Hagiwara:2002,Linde:1998PRD452}
 & \multicolumn{1}{c|}{$\rho_B=0$}
 & \multicolumn{4}{c|}{$\rho_B=\frac{\rho_0}{2}$}
 & \multicolumn{4}{c|}{$\rho_B=\rho_0$} \\
\cline{3-11}
 
& $\mu_B$ (free space)
& $\mu_B^*$
& $\mu^*_{\mathrm{val}}$ & $\mu^*_{\mathrm{sea}}$ & $\mu^*_{\mathrm{orb}}$ & $\mu^*_B$
& $\mu^*_{\mathrm{val}}$ & $\mu^*_{\mathrm{sea}}$ & $\mu^*_{\mathrm{orb}}$ & $\mu^*_B$ \\
\hline

$\mu^*_{\Delta^{++}}$ & $4.5-7.5$
& 5.408
& 5.488 & $-1.143$ & 0.511 & 4.856
& 4.980 & $ -1.040$ & 0.389 & 4.328 \\\hline

$\mu^*_{\Delta^{+}}$ & --
& 2.461
& 2.742 &$ -0.719$ & 0.198 & 2.220
& 2.483 & $-0.655$ & 0.160 & 1.988 \\\hline

$\mu^*_{\Delta^{0}}$ & --
& $-0.487$
& 0.0 & $-0.296$ & $-0.011$ & $-0.411$
& 0.0 & $-0.272$ & $-0.068$ & $-0.340$ \\\hline

$\mu^*_{\Delta^{-}}$ & --
& $-3.434$
& $-2.735$ & 0.125 & $-0.427$ & $-3.038$
& $-2.470$ & 0.109 & $-0.296$ & $-2.657$ \\\hline

$\mu^*_{\Sigma^{*+}}$ & --
& 3.061
& 3.229 & $-0.753$ & 0.334 & 2.810
& 3.012 & $-0.696$ & 0.254 & 2.569 \\\hline

$\mu^*_{\Sigma^{*-}}$ & --
& -2.839
& -2.298 & 0.099 & -0.291 & -2.491
& -2.046 & 0.084 & -0.202 & -2.165 \\\hline

$\mu^*_{\Sigma^{*0}}$ & --
& 0.110
& 0.462 & $-0.326$ & 0.021 & 0.157
& 0.476 & $-0.306$ & 0.025 & 0.196 \\\hline

$\mu^*_{\Xi^{*0}}$ & --
& 0.709
& 0.944 & $ -0.361$ & 0.157 & 0.741
& 0.998 & $-0.349$ & 0.119 & 0.768 \\\hline

$\mu^*_{\Xi^{*-}}$ & --
& $-2.233$
& $-1.865$ & 0.073 & $-0.155$ & $-1.957$
& $-1.634$ & 0.058 & $-0.109$ & $-1.685$ \\\hline

$\mu^*_{\Omega^{-}}$ & $-2.02 \pm 0.05$
& $-1.642$
& $-1.433$ & 0.047 & $-0.019$ & $-1.405$
& $-1.202$ & 0.031 & $-0.016$ & -1.189 \\\hline

\end{tabular}

\caption{Magnetic moments (in units of $\mu_N$) of decuplet baryons at $\rho_B=0$, $\rho_B=\frac{\rho_0}{2}$, and $\rho_B=\rho_0$ for zero magnetic field. The finite density results correspond to $f_s=0.7$ and $I_a=0.3$ whereas $\rho_B=0$ corresponds to $f_s=I_a=0$.}
\label{tab_magmoments1}
\end{table}

\begin{table}[h!]
\centering
\renewcommand{\arraystretch}{1.7}
\setlength{\tabcolsep}{6pt}

\begin{tabular}{|c|c|c|c|c|c|c|c|c|c|}
\hline
 & \multicolumn{1}{c|}{$\rho_B=0$}
 & \multicolumn{4}{c|}{$\rho_B=\frac{\rho_0}{2}$}
 & \multicolumn{4}{c|}{$\rho_B=\rho_0$} \\
\cline{2-10}
 
 & $\mu^*_B$
 & $\mu^*_{\mathrm{val}}$
 & $\mu^*_{\mathrm{sea}}$
 & $\mu^*_{\mathrm{orb}}$
 & $\mu^*_B$
 & $\mu^*_{\mathrm{val}}$
 & $\mu^*_{\mathrm{sea}}$
 & $\mu^*_{\mathrm{orb}}$
 & $\mu^*_B$ \\
\hline

$\mu^*_{\Delta^{++}}$
& 5.864
& 5.847 & $-1.215 $& 0.609 & 5.240
& 5.256 &$ -1.097$ & 0.453 & 4.612 \\\hline

$\mu^*_{\Delta^{+}}$
& 2.657
& 2.920 & $-0.763 $& 0.230 & 2.386
& 2.621 & $-0.690$ & 0.181 & 2.112 \\\hline

$\mu^*_{\Delta^{0}}$
& $-0.548$
& 0.000 & $-0.313$ & $-0.148 $& $-0.461$
& 0.000 & $-0.285$ & $-0.092 $& $-0.377$ \\\hline

$\mu^*_{\Delta^{-}}$
& $-3.750$
& $-2.912$ & 0.136 &$ -0.527$ & $-3.302$
& $-2.607$ & 0.117 & $-0.364$ & $-2.853$ \\\hline

$\mu^*_{\Sigma^{*+}}$
& 3.265
& 3.378 & $-0.793$ & 0.398 & 2.984
& 3.133 & $-0.728$ & 0.296 & 2.701 \\\hline

$\mu^*_{\Sigma^{*-}}$
& $-3.126$
& $-2.477$ & 0.111 & $-0.359$ & $-2.726$
& $-2.178 $ & 0.092 & $-0.248$ & $-2.335$ \\\hline

$\mu^*_{\Sigma^{*0}}$
& 0.068
& 0.446 & $-0.340$ & 0.020 & 0.126
& 0.470 & $-0.317$ & 0.024 & 0.177 \\\hline

$\mu^*_{\Xi^{*0}}$
& 0.679
& 0.900 & $-0.369$ & 0.188 & 0.718
& 0.973 & $-0.356$ & 0.139 & 0.756 \\\hline

$\mu^*_{\Xi^{*-}}$
& $-2.477$
& $-2.036$ & 0.085 & $-0.191$ & $-2.142$
& $-1.756$ & 0.066 & $-0.133$ & $-1.823$ \\\hline

$\mu^*_{\Omega^{-}}$
& $-1.846$
& $-1.606$ & 0.059 & $-0.023 $& $-1.570$
& $-1.325$ & 0.040 & $-0.018$ & $-1.303$ \\\hline

\end{tabular}

\caption{Same as Table~\ref{tab_magmoments1} for magnetic field $eB=3m_\pi^2$.}
\label{tab_magmoments2}

\end{table}
\begin{table}[h!]
\centering
\renewcommand{\arraystretch}{1.7}
\setlength{\tabcolsep}{6pt}

\begin{tabular}{|c|c|c|c|c|c|c|c|c|c|}
\hline
 & \multicolumn{1}{c|}{$\rho_B=0$}
 & \multicolumn{4}{c|}{$\rho_B=\frac{\rho_0}{2}$}
 & \multicolumn{4}{c|}{$\rho_B=\rho_0$} \\
\cline{2-10}
 
 & $\mu^*_B$
 & $\mu^*_{\mathrm{val}}$
 & $\mu^*_{\mathrm{sea}}$
 & $\mu^*_{\mathrm{orb}}$
 & $\mu^*_B$
 & $\mu^*_{\mathrm{val}}$
 & $\mu^*_{\mathrm{sea}}$
 & $\mu^*_{\mathrm{orb}}$
 & $\mu^*_B$ \\
\hline

$\mu^*_{\Delta^{++}}$
& 7.078
& 6.949 & $-1.433$ & 0.991 & 6.507
& 6.200 & $-1.285$ & 0.716 & 5.631 \\\hline

$\mu^*_{\Delta^{+}}$
& 3.173
& 3.469 & $-0.896$ & 0.357 & 2.930
& 3.092 & $-0.805$ & 0.266 & 2.553 \\\hline

$\mu^*_{\Delta^{0}}$
& $-0.725$
& 0.000 & $-0.361$ & $-0.276$ & $-0.637$
& 0.000 & $-0.328$ & $-0.183$ & $-0.511$ \\\hline

$\mu^*_{\Delta^{-}}$
& $-4.616$
& $-3.458$ & 0.172 & $-0.910$ & $-4.195$
& $-3.075$ & 0.147 & $-0.633$ & $-3.561$ \\\hline

$\mu^*_{\Sigma^{*+}}$
& 3.817
& 3.816 & $-0.912$ & 0.649 & 3.553
& 3.520 & $-0.831$ & 0.469 & 3.158 \\\hline

$\mu^*_{\Sigma^{*-}}$
& $-3.922$
& $-3.067$ & 0.150 & $-0.619$ & $-3.536$
& $-2.653$ & 0.122 & $-0.431$ & $-2.961$ \\\hline

$\mu^*_{\Sigma^{*0}}$
& $-0.056$
& 0.369 & $-0.380$ & 0.015 & 0.003
& 0.425 & $-0.353$ & 0.019 & 0.091 \\\hline

$\mu^*_{\Xi^{*0}}$
& 0.603
& 0.722 & $-0.394$ & 0.306 & 0.635
& 0.848 & $-0.377$ & 0.221 & 0.692 \\\hline

$\mu^*_{\Xi^{*-}}$
& $-3.168$
& $-2.632$ & 0.125 & $-0.327$ & $-2.834$
& $-2.212$ & 0.097 & $-0.229 $& $-2.344$ \\\hline

$\mu^*_{\Omega^{-}}$
& $-2.413$
& $-2.208$ & 0.101 & $-0.036$ & $-2.143$
& $-1.787$ & 0.072 & $-0.026$ & $-1.742$ \\\hline

\end{tabular}

\caption{Same as Table~\ref{tab_magmoments1} for magnetic field $eB=5m_\pi^2$.}
\label{tab_magmoments3}
\end{table}

\section{Summary}   
\label{sec:conclusion}
To summarize, in the present work we investigated the in-medium effective masses and magnetic moments of the decuplet baryon in isospin asymmetric magnetized strange matter at finite temperature, within the CQMF framework including Dirac sea contributions. We observe that the magnitudes of the scalar condensates increase with increasing magnetic field strength due to the contribution from the Dirac sea, thereby exhibiting the phenomenon of magnetic catalysis. The $\sigma$ field shows the strongest magnetic sensitivity, whereas the $\zeta$ field is more sensitive to strangeness fraction of the medium. The effective masses of all decuplet baryons increase with magnetic field strength, reflecting the enhancement of scalar condensates, while the finite baryon density reduces the masses consistent with the previous studies on the masses of baryons. The isospin asymmetry  effects are most pronounced for baryons composed of entirely of light quarks through $\delta$ meson field which induces the charge splitting within each multiplet whereas the number of strange quarks determines the effects of strangeness fraction in baryons. Therefore, $\Delta$ baryons show significant change with $I_a$ whereas $\Omega^-$ baryon shows more pronounced change with $f_s$. 
The magnetic moments of the decuplet baryons are evaluated using the in-medium quark and baryon masses as inputs to the chiral constituent quark model. The total magnetic moments receive contributions from valence quarks, sea quarks, and orbital angular momentum of the quark sea, with the valence contribution dominating in all cases. The magnitude of total magnetic moments of the baryons are enhanced in the strange medium as a function of magnetic field. The total magnetic moments of most decuplet baryons decrease in magnitude with increasing baryon density. However, the $\Sigma^{*0}$ and $\Xi^{*0}$ baryons display the opposite behavior, exhibiting an increase in magnitude as the density increases. 
These results may serve as important theoretical inputs for future investigations of strongly magnetized dense matter, including magnetar interiors and the post-merger phase of binary neutron star collisions, where high density, strong magnetic fields, and strangeness coexist simultaneously. This study is also motivated by heavy-ion collision experiments at RHIC (Relativistic Heavy Ion Collider) and the LHC (Large Hadron Collider), where extremely strong transient magnetic fields and hot dense matter are produced~\cite{Kharzeev:2008npa,Skokov:2009mag,Voronyuk:2011prc,Bzdak:2012plb,Deng:2012prc}. 
\newpage
 \section*{Acknowledgment}
H.D. would like to thank the Science and Engineering Research Board, Anusandhan-National
Research Foundation (ANRF), Government of India under the SERB-POWER Fellowship
scheme (Ref No. SPF/2023/000116). A.K. sincerely acknowledge Anusandhan-National Research Foundation (ANRF), Government of India for funding of the research project under the Science and Engineering
Research Board-Core Research Grant (SERB-CRG) scheme (File No. CRG/2023/000557) for financial support.

\bibliographystyle{elsarticle-num} 
\bibliography{Ref}

@article{Parui:2022hepph,
  author = {Parui, Pallabi and De, Sourodeep and Kumar, Ankit and Mishra, Amruta},
  journal = {Phys. Rev. D},
  volume = {$\textbf{106}$},
  pages = {114033},
  year = {2022},
  publisher = {American Physical Society}
}

@article{De:2022hepph,
  author = {De, Sourodeep and Parui, Pallabi and Mishra, Amruta},
  journal = {Phys. Rev. C},
  volume = {$\textbf{107}$},
  pages = {065204},
  numpages = {21}
}

@article{Parui:2022openbottom,
  author    = {De, Sourodeep and Parui, Pallabi and Mishra, Amruta},
  journal   = {Int. J. Mod. Phys. E},
  volume    = {$\textbf{31}$},
  pages     = {2250106},
  year      = {2022},
}

@article{FloresMendieta:2009,
  author  = {Flores Mendieta, R.},
  journal = {Phys. Rev. D},
  volume  = {$\textbf{80}$},
  pages   = {094014},
  year    = {2009}
}

@article{Geng:2009,
  author  = {Geng, L. S. and Camalich, J. M. and Vacas, M. J. V.},
  journal = {Phys. Rev. D},
  volume  = {$\textbf{80}$},
  pages   = {034027},
  year    = {2009}
}

@article{Lee:2005ds,
  author  = {Lee, F. X. and Kelly, R. and Zhou, L. and Wilcox, W.},
  journal = {Phys. Lett. B},
  volume  =  {$\textbf{627}$},
  pages   = {71},
  year    = {2005}
}

@article{Broderick:2000,
  author  = {Broderick, A. and Prakash, M. and Lattimer, J. M.},
  journal = {Astrophys. J.},
  volume  = {$\textbf{537}$},
  pages   = {351},
  year    = {2000}
}

@article{Wei:2006,
  author  = {Wei, F. X. and \textit{et al.}},
  journal = {J. Phys. G: Nucl. Part. Phys.},
  volume  = {$\textbf{32}$},
  pages   = {47},
  year    = {2006}
}

@article{Hagiwara:2002,
  author    = {Hagiwara, K. and \textit{et al.}},
  journal   = {Phys. Rev. D},
  volume    = {$\textbf{66}$},
  pages     = {010001},
  year      = {2002}
}

@article{Mao:2003,
  author  = {Mao, G.-J. and Iwamoto, A. and Li, Z.-X.},
  journal = {Chin. J. Astrophys.},
  volume  = {$\textbf{3}$},
  pages   = {359},
  year    = {2003}
}

@article{Dahiya:2023izc,
  author    = {Dahiya, H. and Dutt, S. and Kumar, A. and Randhawa, M.},
  journal   = {Eur. Phys. J. Plus},
  volume    = {$\textbf{138}$},
  number    = {5},
  year      = {2023}
}

@article{Voronyuk:2011prc,
  author  = {Voronyuk, V. and \textit{et al.}},
  journal = {Phys. Rev. C},
  volume  = {$\textbf{83}$},
  pages   = {054911},
  year    = {2011}
}

@article{Bzdak:2012plb,
  author  = {Bzdak, A. and Skokov, V.},
  journal = {Phys. Lett. B},
  volume  = {$\textbf{710}$},
  pages   = {171},
  year    = {2012}
}

@article{Deng:2012prc,
  author  = {Deng, W.-T. and Huang, X.-G.},
  journal = {Phys. Rev. C},
  volume  = {$\textbf{85}$},
  pages   = {044907},
  year    = {2012}
}

@article{Bloczynski:2013plb,
  author  = {Bloczynski, J. and Huang, X.-G. and Zhang, X. and Liao, J.},
  journal = {Phys. Lett. B},
  volume  = {$\textbf{718}$},
  pages   = {1529},
  year    = {2013}
}

@article{McLerran:2014npa,
  author  = {McLerran, L. and Skokov, V.},
  journal = {Nucl. Phys. A},
  volume  = {$\textbf{929}$},
  pages   = {184},
  year    = {2014}
}

@article{Kharzeev:2016ppnp,
  author  = {Kharzeev, D. E. and Liao, J. and Voloshin, S. A. and Wang, G.},
  journal = {Prog. Part. Nucl. Phys.},
  volume  = {$\textbf{88}$},
  pages   = {1},
  year    = {2016}
}

@article{Kharzeev:2021natrev,
  author  = {Kharzeev, D. E. and Liao, J.},
  journal = {Nat. Rev. Phys.},
  volume  = {$\textbf{3}$},
  pages   = {55},
  year    = {2021}
}

@article{Fukushima:2019ppnp,
  author  = {Fukushima, K.},
  journal = {Prog. Part. Nucl. Phys.},
  volume  = {$\textbf{107}$},
  pages   = {167},
  year    = {2019}
}

@article{Shovkovy:2022ws,
  author       = {Shovkovy, Igor A.},
  journal      = {Particles},
  volume       = {$\textbf{5}$},
  pages        = {442},
  year         = {2022}
}

@article{Gusynin:1996npb,
  author  = {Gusynin, V. P. and Miransky, V. A. and Shovkovy, I. A.},
  journal = {Nucl. Phys. B},
  volume  = {$\textbf{462}$},
  pages   = {249},
  year    = {1996}
}

@article{Bali:2012prd,
  author  = {Bali, G. S. and \textit{et al.}},
  journal = {Phys. Rev. D},
  volume  = {$\textbf{86}$},
  pages   = {071502},
  year    = {2012}
}

@article{Bali:2013jhep2,
  author  = {Bali, G. S. and Bruckmann, F. and Endrodi, G. and Gruber, F. and Schafer, A.},
  journal = {JHEP},
  volume  = {$\textbf{04}$},
  pages   = {130},
  year    = {2013}
}

@article{Kharzeev:2011prl,
  author  = {Kharzeev, D. E. and Son, D. T.},
  journal = {Phys. Rev. Lett.},
  volume  = {$\textbf{106}$},
  pages   = {062301},
  year    = {2011}
}

@article{Fayazbakhsh:2014prd,
  author  = {Fayazbakhsh, S. and \textit{et al.}},
  journal = {Phys. Rev. D},
  volume  = {$\textbf{90}$},
  pages   = {105030},
  year    = {2014}
}

@article{Ferrer:2014prd,
  author  = {Ferrer, E. J. and \textit{et al.}},
  journal = {Phys. Rev. D},
  volume  = {$\textbf{89}$},
  pages   = {085034},
  year    = {2014}
}

@article{Huang:2023prc,
  author  = {Huang, A. and \textit{et al.}},
  journal = {Phys. Rev. C},
  volume  = {$\textbf{107}$},
  pages   = {034901},
  year    = {2023}
}

@article{Li:2020arnps,
  author  = {Li, W. and Wang, G.},
  journal = {Annu. Rev. Nucl. Part. Sci.},
  volume  = {$\textbf{70}$},
  pages   = {293},
  year    = {2020}
}

@article{Cheng:1995,
  author  = {Cheng, T. P. and Li, L. F.},
  journal = {Phys. Rev. Lett.},
  volume  = {$\textbf{74}$},
  pages   = {2872},
  year    = {1995}
}

@article{Cheng:1998PRD,
  author  = {Cheng, T. P. and Li, L. F.},
  journal = {Phys. Rev. D},
  volume  = {$\textbf{57}$},
  pages   = {344},
  year    = {1998}
}

@article{Linde:1998PRD452,
  author  = {Linde, J. and Ohlsson, T. and Snellman, H.},
  journal = {Phys. Rev. D},
  volume  = {$\textbf{57}$},
  pages   = {452},
  year    = {1998}
}

@article{Song:1997,
  author  = {Song, X. and McCarthy, J. S. and Weber, H. J.},
  journal = {Phys. Rev. D},
  volume  = {$\textbf{55}$},
  pages   = {2624},
  year    = {1997}
}

@article{Song:1998,
  author  = {Song, X.},
  journal = {Phys. Rev. D},
  volume  = {$\textbf{57}$},
  pages   = {4114},
  year    = {1998}
}

@article{Dahiya:2004IJMPA,
  author  = {Dahiya, H. and Gupta, M.},
  journal = {Int. J. Mod. Phys. A},
  volume  = {$\textbf{19}$},
  pages   = {5027},
  year    = {2004}
}

@article{Dahiya:2006IJMPA,
  author  = {Dahiya, H. and Gupta, M. and Rana, J. M. S.},
  journal = {Int. J. Mod. Phys. A},
  volume  = {$\textbf{21}$},
  pages   = {4255},
  year    = {2006}
}

@article{Huang:2016rpp,
  author  = {Huang, X.-G.},
  journal = {Rep. Prog. Phys.},
  volume  = {$\textbf{79}$},
  pages   = {076302},
  year    = {2016}
}

@article{pwang2003,
  author = {Wang, P. and Lyubovitskij, V. E. and Gutsche, Th. and Faessler, Amand},
  journal = {Phys. Rev. C},
  volume = {$\textbf{67}$},
  pages = {015210},
  numpages = {10},
  year = {2003}
}

@article{Tuchin:2016prc,
  author  = {Tuchin, K.},
  journal = {Phys. Rev. C},
  volume  = {$\textbf{93}$},
  pages   = {014905},
  year    = {2016}
}

@article{Chen:2021npa,
  author  = {Chen, Y. and Sheng, X.-L. and Ma, G.-L.},
  journal = {Nucl. Phys. A},
  volume  = {$\textbf{1011}$},
  pages   = {122199},
  year    = {2021}
}

@article{Fraga:2024prd,
  author  = {Fraga, E. S. and Palhares, L. F. and Villavicencio, C.},
  journal = {Phys. Rev. D},
  volume  = {$\textbf{109}$},
  number  = {11},
  pages   = {116018},
  year    = {2024}
}

@article{Ping:2001ctp,
  author  = {Ping, W. and Zong-Ye, Z. and You-Wen, Y.},
  journal = {Commun. Theor. Phys.},
  volume  = {$\textbf{36}$},
  pages   = {71},
  year    = {2001}
}

@book{Landau:1965,
  author    = {Landau, L. D. and ter Haar, D.},
  title     = {{Collected Papers of L. D. Landau}},
  publisher = {Pergamon Press},
  year      = {1965}
}

@article{Singh:2017cpc,
  author  = {Singh, H. and Kumar, A. and Dahiya, H.},
  journal = {Chinese Phys. C},
  volume  = {$\textbf{41}$},
  pages   = {094104},
  year    = {2017}
}

@article{Singh:2019epjp,
  author  = {Singh, H. and Kumar, A. and Dahiya, H.},
  journal = {Eur. Phys. J. Plus},
  volume  = {$\textbf{134}$},
  pages   = {128},
  year    = {2019}
}

@article{Ryu:2010prc,
  author  = {Ryu, C. Y. and Kim, K. S.},
  journal = {Phys. Rev. C},
  volume  = {$\textbf{82}$},
  pages   = {025804},
  year    = {2010}
}

@article{Rezaei:2018ijmpe,
  author  = {Rezaei, Z.},
  journal = {Int. J. Mod. Phys. E},
  volume  = {$\textbf{27}$},
  pages   = {1850011},
  year    = {2018}
}

@article{Singh:2018epja,
  author  = {Singh, H. and Kumar, A. and Dahiya, H.},
  journal = {Eur. Phys. J. A},
  volume  = {$\textbf{54}$},
  pages   = {120},
  year    = {2018}
}

@article{Singh:2020epjp,
  author  = {Singh, H. and Kumar, A. and Dahiya, H.},
  journal = {Eur. Phys. J. Plus},
  volume  = {$\textbf{135}$},
  pages   = {422},
  year    = {2020}
}

@article{Wang:2001npa,
  author  = {Wang, P. and \textit{et al.}},
  journal = {Nucl. Phys. A},
  volume  = {$\textbf{688}$},
  pages   = {791},
  year    = {2001}
}

@article{Tsushima:2022ptep,
  author  = {Tsushima, K.},
  journal = {Prog. Theor. Exp. Phys.},
  volume  = {$\textbf{2022}$},
  pages   = {043D02},
  year    = {2022}
}

@article{Broderick:2002plb,
  author  = {Broderick, A. E. and Prakash, M. and Lattimer, J. M.},
  journal = {Phys. Lett. B},
  volume  = {$\textbf{531}$},
  pages   = {167},
  year    = {2002}
}

@article{Menezes:2009prc,
  author  = {Menezes, D. P. and \textit{et al.}},
  journal = {Phys. Rev. C},
  volume  = {$\textbf{79}$},
  pages   = {035807},
  year    = {2009}
}

@article{Moreira:2021epja,
  author  = {Moreira, J. and Costa, P. and Restrepo, T. E.},
  journal = {Eur. Phys. J. A},
  volume  = {$\textbf{57}$},
  pages   = {123},
  year    = {2021}
}

@article{Wang:2022prd,
  author  = {Wang, Y. and Wen, X.-J.},
  journal = {Phys. Rev. D},
  volume  = {$\textbf{105}$},
  pages   = {074034},
  year    = {2022}
}

@article{Yue:2008prc,
  author  = {Yue, P. and Shen, H.},
  journal = {Phys. Rev. C},
  volume  = {$\textbf{77}$},
  pages   = {045804},
  year    = {2008}
}

@article{Dexheimer:2012epja,
  author  = {Dexheimer, V. and Negreiros, R. and Schramm, S.},
  journal = {Eur. Phys. J. A},
  volume  = {$\textbf{48}$},
  pages   = {189},
  year    = {2012}
}

@article{Chahal:2023prc,
  author  = {Chahal, N. and Dutt, S. and Kumar, A.},
  journal = {Phys. Rev. C},
  volume  = {$\textbf{107}$},
  pages   = {045203},
  year    = {2023}
}

@article{Chu:2018plb,
  author  = {Chu, P. C. and \textit{et al.}},
  journal = {Phys. Lett. B},
  volume  = {$\textbf{778}$},
  pages   = {447},
  year    = {2018}
}

@article{Sinha:2013npa,
  author  = {Sinha, M. and Mukhopadhyay, B. and Sedrakian, A.},
  journal = {Nucl. Phys. A},
  volume  = {$\textbf{898}$},
  pages   = {43},
  year    = {2013}
}

@article{Gridhar:2015prd,
  author  = {Gridhar, A. and Dahiya, H. and Randhawa, M.},
  journal = {Phys. Rev. D},
  volume  = {$\textbf{92}$},
  pages   = {033012},
  year    = {2015}
}

@article{NSharma2010,
  author  = {Sharma, N. and \textit{et al.}},
  journal = {Phys. Rev. D},
  volume  = {$\textbf{81}$},
  pages   = {073001},
  year    = {2010}
}

@article{Kumari:2020mci,
  author  = {Kumari, M. and Kumar, A.},
  journal = {Eur. Phys. J. Plus},
  volume  = {$\textbf{136}$},
  pages   = {19},
  year    = {2021}
}

@article{Tsushima:1998npa,
  author  = {Tsushima, K. and Saito, K. and Haidenbauer, J. and Thomas, A. W.},
  journal = {Nucl. Phys. A},
  volume  = {$\textbf{630}$},
  pages   = {691},
  year    = {1998}
}

@article{Kharzeev:2008npa,
  author  = {Kharzeev, D. E. and McLerran, L. D. and Warringa, H. J.},
  journal = {Nucl. Phys. A},
  volume  = {$\textbf{803}$},
  pages   = {227},
  year    = {2008}
}

@article{Fukushima:2008prd,
  author  = {Fukushima, K. and Kharzeev, D. E. and Warringa, H. J.},
  journal = {Phys. Rev. D},
  volume  = {$\textbf{78}$},
  pages   = {074033},
  year    = {2008}
}

@article{Skokov:2009mag,
  author  = {Skokov, V. and Illarionov, A. and Toneev, V.},
  journal = {Int. J. Mod. Phys. A},
  volume  = {$\textbf{24}$},
  pages   = {5925},
  year    = {2009}
}

@article{Kumar:2023owb,
  author  = {Kumar, A. and Dutt, S. and Dahiya, H.},
  journal = {Eur. Phys. J. A},
  volume  = {$\textbf{60}$},
  pages   = {4},
  year    = {2024}
}

@article{Menezes:2009prc2,
  author  = {Menezes, D. P. and \textit{et al.}},
  journal = {Phys. Rev. C},
  volume  = {$\textbf{80}$},
  pages   = {065805},
  year    = {2009}
}

@article{Haber:2014prd,
  author  = {Haber, A. and Preis, F. and Schmitt, A.},
  journal = {Phys. Rev. D},
  volume  = {$\textbf{90}$},
  pages   = {125036},
  year    = {2014}
}

@article{Mukherjee:2018prd,
  author  = {Mukherjee, A. and Ghosh, S. and Mandal, M. and \textit{et al.}},
  journal = {Phys. Rev. D},
  volume  = {$\textbf{98}$},
  pages   = {056024},
  year    = {2018}
}

@article{Aguirre:2016epja,
  author  = {Aguirre, R. M. and De Paoli, A. L.},
  journal = {Eur. Phys. J. A},
  volume  = {$\textbf{52}$},
  pages   = {343},
  year    = {2016}
}

@article{Aguirre:2019prc,
  author  = {Aguirre, R. M.},
  journal = {Phys. Rev. C},
  volume  = {$\textbf{100}$},
  pages   = {065203},
  year    = {2019}
}

@article{Rabhi:2011prc,
  author  = {Rabhi, A. and Panda, P. K. and Providencia, C.},
  journal = {Phys. Rev. C},
  volume  = {$\textbf{84}$},
  pages   = {035803},
  year    = {2011}
}

@article{Bali:2012jhep,
  author  = {Bali, G. S. and \textit{et al.}},
  journal = {JHEP},
  volume  = {$\textbf{02}$},
  pages   = {044},
  year    = {2012}
}

@article{Bernard:1995ijmpe,
  author  = {Bernard, V. and Kaiser, N. and Meissner, U.-G.},
  journal = {Int. J. Mod. Phys. E},
  volume  = {$\textbf{4}$},
  pages   = {193},
  year    = {1995}
}

@article{Papazoglou:1999prc,
  author  = {Papazoglou, P. and \textit{et al.}},
  journal = {Phys. Rev. C},
  volume  = {$\textbf{59}$},
  pages   = {411},
  year    = {1999}
}

@article{Manohar:1984georgi,
  author  = {Manohar, A. and Georgi, H.},
  journal = {Nucl. Phys. B},
  volume  = {$\textbf{234}$},
  pages   = {189},
  year    = {1984}
}

@article{Cheng:1997prl,
  author  = {Cheng, T. P. and Li, L.-F.},
  journal = {Phys. Rev. Lett.},
  volume  = {$\textbf{80}$},
  pages   = {2789},
  year    = {1998}
}

@article{Dahiya:2001prd,
  author  = {Dahiya, Harleen and Gupta, M.},
  journal = {Phys. Rev. D},
  volume  = {$\textbf{64}$},
  pages   = {014013},
  year    = {2001}
}

@article{Agasian:2008plb,
  author  = {Agasian, N. O. and Fedorov, S. M.},
  journal = {Phys. Lett. B},
  volume  = {$\textbf{663}$},
  pages   = {445},
  year    = {2008}
}

@article{Mishra2024Charmonium,
  author  = {Mishra, A. and Kumar, A. and Misra, S. P.},
  journal = {Phys. Rev. D},
  volume  = {$\textbf{110}$},
  pages   = {014003},
  year    = {2024}
}

@article{Miransky_2002,
  author  = {Miransky, V. A. and Shovkovy, I. A.},
  journal = {Phys. Rev. D},
  volume  = {$\textbf{66}$},
  pages  = {045006},
  year    = {2002}
}

@article{Constantinescu:1972ie,
  author  = {Constantinescu, D. H.},
  journal = {Nucl. Phys. B},
  volume  = {$\textbf{36}$},
  pages   = {121},
  year    = {1972}
}

@article{Dastidar:2026beh,
  author  = {Dastidar, Utsa and Kumar, Arvind and Dahiya, Harleen and Dutt, S.},
  volume  = {arXiv:2603.23586},
  year    = {2026}
}

@article{PhysRevD.52.4099,
  author  = {Sahoo, R. K. and Panda, A. R. and Nath, A.},
  journal = {Phys. Rev. D},
  volume  = {$\textbf{52}$},
  pages   = {4099},
  year    = {1995}
}

@article{PhysRevD.79.094025,
  author  = {Ledwig, T. and Silva, A. and Vanderhaeghen, M.},
  journal = {Phys. Rev. D},
  volume  = {$\textbf{79}$},
  pages   = {094025},
  year    = {2009}
}

@article{PhysRevD.105.096002,
  author  = {Fu, D. and Sun, B.-D. and Dong, Y.},
  journal = {Phys. Rev. D},
  volume  = {$\textbf{105}$},
  pages   = {096002},
  year    = {2022}
}

@article{Wen2025EPJC,
  author  = {Wen, L.-Z. and Chen, Y.-K. and Meng, L. and Zhu, S.-L.},
  journal = {Eur. Phys. J. C},
  volume  = {$\textbf{85}$},
  pages   = {1210},
  year    = {2025}
}

@article{PhysRevD.65.073017,
  author  = {Buchmann, A. J. and Henley, E. M.},
  journal = {Phys. Rev. D},
  volume  = {$\textbf{65}$},
  pages   = {073017},
  year    = {2002}
}

@article{PhysRevD.92.014038,
  author  = {Taya, H.},
  journal = {Phys. Rev. D},
  volume  = {$\textbf{92}$},
  pages   = {014038},
  year    = {2015}
}

@article{PhysRevD.28.2881,
  author  = {Fajfer, S. and Oakes, R. J.},
  journal = {Phys. Rev. D},
  volume  = {$\textbf{28}$},
  pages   = {2881},
  year    = {1983}
}

@article{PhysRevD.44.1962,
  author  = {Bosshard, A. and \textit{et al.}},
  journal = {Phys. Rev. D},
  volume  = {$\textbf{44}$},
  pages   = {1962},
  year    = {1991}
}

@article{PhysRevLett.67.804,
  author  = {Diehl, H. T. and \textit{et al.}},
  journal = {Phys. Rev. Lett.},
  volume  = {$\textbf{67}$},
  pages   = {804},
  year    = {1991}
}

@article{PhysRevLett.74.3732,
  author  = {Wallace, N. B. and \textit{et al.}},
  journal = {Phys. Rev. Lett.},
  volume  = {$\textbf{74}$},
  pages   = {3732},
  year    = {1995}
}

@article{PhysRevLett.89.272001,
  author  = {Kotulla, M. and \textit{et al.}},
  journal = {Phys. Rev. Lett.},
  volume  = {$\textbf{89}$},
  pages   = {272001},
  year    = {2002}
}

@article{Leinweber1992,
  author  = {Leinweber, D. B. and Draper, T. and Woloshyn, R. M.},
  journal = {Phys. Rev. D},
  volume  = {$\textbf{46}$},
  pages   = {3067},
  year    = {1992}
}

@article{Alexandrou2009,
  author  = {Alexandrou, C. and Korzec, T. and Koutsou, G. and Leontiou, T.},
  journal = {Phys. Rev. D},
  volume  = {$\textbf{79}$},
  pages   = {014507},
  year    = {2009}
}

@article{Kaur2016EPJA,
  author  = {Kaur, A. and Upadhyay, A.},
  journal = {Eur. Phys. J. A},
  volume  = {$\textbf{52}$},
  pages   = {105},
  year    = {2016}
}

@article{Savage2002,
  author  = {Savage, M. J.},
  journal = {Nucl. Phys. A},
  volume  = {$\textbf{700}$},
  pages   = {359},
  year    = {2002}
}

@article{PhysRevD.48.4478,
  author  = {Schlumpf, F.},
  journal = {Phys. Rev. D},
  volume  = {$\textbf{48}$},
  pages   = {4478},
  year    = {1993}
}

@article{PhysRevD.57.1801,
  author  = {Lee, F. X.},
  journal = {Phys. Rev. D},
  volume  = {$\textbf{57}$},
  pages   = {1801},
  year    = {1998}
}

@article{PhysRevD.67.114015,
  author  = {Dahiya, H. and Gupta, M.},
  journal = {Phys. Rev. D},
  volume  = {$\textbf{67}$},
  pages   = {114015},
  year    = {2003}
}

@article{Dahiya:2019owy,
  author  = {Dahiya, H.},
  journal = {JPS Conf. Proc.},
  volume  = {$\textbf{26}$},
  pages   = {021019},
  year    = {2019}
}

@article{Kumar:2019cpc,
  author  = {Kumar, R. and Kumar, A.},
  journal = {Chinese Phys. C},
  volume  = {$\textbf{43}$},
  pages   = {124109},
  year    = {2019}
}

@article{Kumar:2019epjc,
  author  = {Kumar, R. and Kumar, A.},
  journal = {Eur. Phys. J. C},
  volume  = {$\textbf{79}$},
  pages   = {403},
  year    = {2019}
}

@article{Barik:1985prd,
  author  = {Barik, N. and Dash, B. K.},
  journal = {Phys. Rev. D},
  volume  = {$\textbf{31}$},
  pages   = {7},
  year    = {1985}
}

@article{Barik:2013prc,
  author  = {Barik, N. and Mishra, A. and Dash, B. K.},
  journal = {Phys. Rev. C},
  volume  = {$\textbf{88}$},
  pages   = {015206},
  year    = {2013}
}

@article{Mukherjee:2018ebw,
	author = {Mukherjee, Arghya and Ghosh, Snigdha and Mandal, Mahatsab and Sarkar, Sourav and Roy, Pradip},
	journal = {Phys. Rev. D},
	volume = {$\textbf{98}$},
	pages = {056024},
	year = {2018}
}

\section*{Appendix}
\begin{sidewaystable}[h!]
\centering
\renewcommand{\arraystretch}{1}
\begin{tabular}{|c|c|c|c|c|}
\hline
Baryon & $\Delta u$ & $\Delta d$ & $\Delta s$ & $\Delta c$ \\

\hline

$\Delta^{++}(uuu)$ 

& $3 - a\left(6 + 3\alpha^2 + \beta^2 + \frac{\zeta^2}{8} + \frac{5\gamma^2}{16}\right)$

& $-3a$

& $-3a\alpha^2$

& $-3a\gamma^2$ \\

$\Delta^{+}(uud)$ 

& $2 - a\left(5 + 2\alpha^2 + \frac{2\beta^2}{3} + \frac{\zeta^2}{12} + \frac{17\gamma^2}{8}\right)$

& $1 - a\left(4 + \alpha^2 + \frac{\beta^2}{3} + \frac{\zeta^2}{12} + \frac{17\gamma^2}{8}\right)$

& $-3a\alpha^2$

& $-3a\gamma^2$ \\

$\Delta^{0}(udd)$ 

& $1 - a\left(4 + \alpha^2 + \frac{\beta^2}{3} + \frac{\zeta^2}{12} + \frac{17\gamma^2}{8}\right)$

& $2 - a\left(5 + 2\alpha^2 + \frac{2\beta^2}{3} + \frac{\zeta^2}{12} + \frac{17\gamma^2}{8}\right)$

& $-3a\alpha^2$

& $-3a\gamma^2$ \\

$\Delta^{-}(ddd)$ 

& $-3a$

& $3 - a\left(6 + 3\alpha^2 + \beta^2 + \frac{\zeta^2}{8} + \frac{5\gamma^2}{16}\right)$

& $-3a\alpha^2$

& $-3a\gamma^2$ \\

$\Sigma^{*+}(uus)$ 

& $2 - a\left(3 + 2\alpha^2 + \frac{\zeta^2}{24} + \frac{17\gamma^2}{16}\right)$

& $-a\left(2 + \alpha^2\right)$

& $1 - a\left(4\alpha^2 + \frac{4\beta^2}{3} + \frac{\zeta^2}{24} + \frac{17\gamma^2}{16}\right)$

& $-3a\gamma^2$ \\

$\Sigma^{*0}(uds)$ 

& $1 - a\left(2 + \alpha^2\right)$

& $1 - a\left(2 + \alpha^2\right)$

& $1 - a\left(4\alpha^2 + \frac{4\beta^2}{3} + \frac{\zeta^2}{24} + \frac{17\gamma^2}{16}\right)$

& $-3a\gamma^2$ \\

$\Sigma^{*-}(dds)$ 

& $-a\left(2 + \alpha^2\right)$

& $2 - a\left(3 + 2\alpha^2 + \frac{\zeta^2}{24} + \frac{17\gamma^2}{16}\right)$

& $1 - a\left(4\alpha^2 + \frac{4\beta^2}{3} + \frac{\zeta^2}{24} + \frac{17\gamma^2}{16}\right)$

& $-3a\gamma^2$ \\

$\Xi^{*0}(uss)$ 

& $1 - a\left(2 + \alpha^2\right)$

& $-a\left(1 + 2\alpha^2\right)$

& $2 - a\left(5\alpha^2 + \frac{8\beta^2}{3} + \frac{\zeta^2}{12} + \frac{17\gamma^2}{8}\right)$

& $-3a\gamma^2$ \\

$\Xi^{*-}(dss)$ 

& $-a\left(1 + 2\alpha^2\right)$

& $1 - a\left(2 + \alpha^2\right)$

& $2 - a\left(5\alpha^2 + \frac{8\beta^2}{3} + \frac{\zeta^2}{12} + \frac{17\gamma^2}{8}\right)$

& $-3a\gamma^2$ \\

$\Omega^{-}(sss)$ 

& $-3a\alpha^2$

& $-3a\alpha^2$

& $3 - a\left(6\alpha^2 + 4\beta^2 + \frac{\zeta^2}{8} + \frac{51\gamma^2}{12}\right)$

& $-3a\gamma^2$ \\

\hline

\end{tabular}

\caption{Quark spin polarizations ($\Delta u, \Delta d, \Delta s, \Delta c$) for spin-$\frac{3}{2}^{+}$ decuplet baryons in the $\chi$CQM framework~\cite{Dahiya:2023izc}.}
\label{tab_magmoments_eqn}
\end{sidewaystable}
\end{document}